\documentclass[14pt]{article}
\usepackage{moreverb}
\usepackage{mathtools}
\usepackage{amsfonts,latexsym}
\usepackage{amsfonts,latexsym}
\usepackage{epsfig}
\usepackage{amsmath}
\usepackage{tcolorbox}
\usepackage{marginnote}
\tcbuselibrary{skins,xparse,breakable}
\begin{document}
\title{Domain Decomposition in space-time for 4D-Var Data Assimilation problem:\\ a case study on the Regional Ocean Modeling System}
\author{R. Cacciapuoti$^a$, L. D'Amore$^a$ and A. M. Moore$^b$. }

\date{$^a$  University of Naples, \textit{Federico II}, Naples (IT).\\
$^b$ Ocean Sciences Department Institute of Marine Sciences, UC Santa Cruz, (CA-USA).}

\maketitle
\noindent 

\section{ROMS}\label{ROMS}

\begin{figure}[!ht]
  \includegraphics[width=1.1\textwidth]{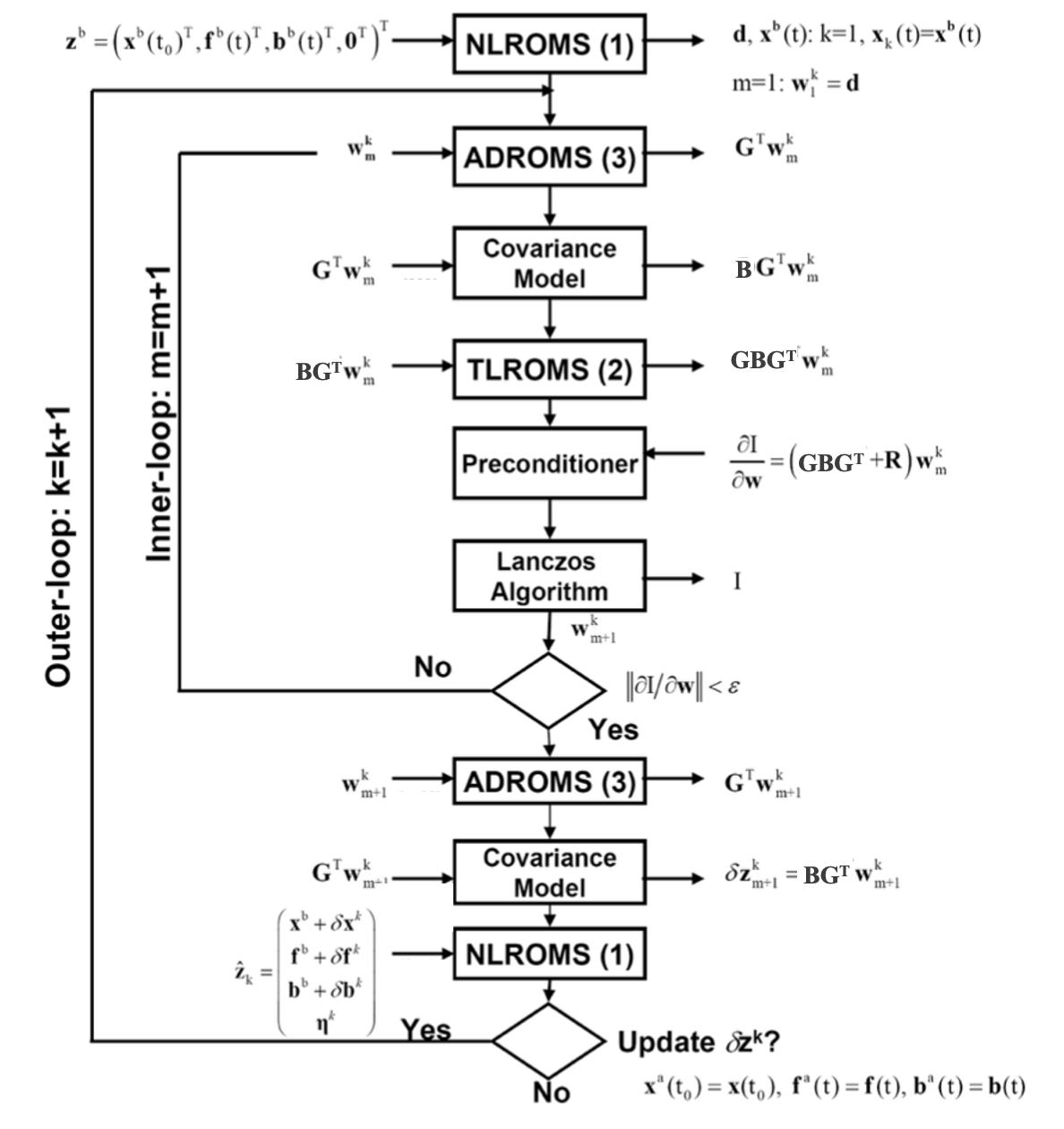}
  \caption{The flow chart illustrating RBL4DVAR algorithm.}
\label{schema_duale}
\end{figure}

Regional Ocean Modeling System (ROMS) is an open-source, mature numerical framework used by both the scientific and operational communities to study ocean dynamics over 3D spatial domain and time interval. \\
ROMS supports different 4D-Var data assimilation (DA) methodologies
such as incremental strong constraint 4D-Var (IS4D-Var) and dual formulation 4D-Var using restricted B-preconditioned Lanczos formulation of the conjugate gradient method (RBL4DVAR) \cite{RBL4DVAR}.
IS4D-Var and RBL4DVAR search best circulation estimate in space spanned by control vector and observations, respectively.
\noindent IS4D-Var and RBL4DVAR algorithm consist of two nested loop, the outer-loop involves the module (1), namely nonlinear ROMS (NLROM) solving ROMS  equations, the inner-loop involves modules (2)-(3), namely tangent linear approximation of ROMS (TLROMS) and adjoint model of ROMS (ADROMS); TLROMS and ADROMS are used for minimizing 4D-Var functional \cite{ROMS_moore} (see Figures \ref{Fig_flow_chart}-\ref{schema_duale}).\\
NLROMS is a three-dimensional, free-surface, terrain-following ocean model that solves the Reynolds-averaged Navier-Stokes equations using the hydrostatic vertical momentum balance and Boussinesq approximation.\\
\noindent
NLROMS computes
\begin{equation}\label{NLROMS}
x^{ROMS}(t_l) = M_{l-1,l}(x(t_{l-1}), f(t_l), b(t_l))
\end{equation}
with the state-vector $x^{ROMS}(t_l) = (T, S,\varsigma , u, v)^T$, temperature $T$, salinity $S$, $(x,y)$ components of vector velocity $u, v$, sea surface displacement $\varsigma$.  $M_{l-1,l}$ represents nonlinear ROMS acting on $x^{ROMS}(t_{l-1})$, and subject to forcing $f(t_l)$, and boundary conditions $b(t_l)$ during the time interval $[t_{l-1},t_l]$. \\

\noindent Minimization of the 4D-Var functional: 
\begin{equation}\label{funzionale}
\mathbf{J}^{ROMS}(\delta z) = \frac{1}{2}\delta z \mathbf{B}^{-1}\delta z + \frac{1}{2} (G\delta z - \mathbf{d})^T \mathbf{R}^{-1}(G\delta z - \mathbf{d})
\end{equation}

\noindent where $\delta z$ are the control variable increments,  $\mathbf{d}$ is vector of innovations, $G = (...,H_l^T ,...)^T$, where $H_l$ is the observation matrix; $\mathbf{R}$ is observation error covariance matrix and $\mathbf{B}$ is covariance matrix of model error, is computed in the inner-loop in Figure \ref{Fig_flow_chart}. \\
\noindent Analysis increment, $\delta z^a$, that minimizes 4D-Var function in (\ref{funzionale}) corresponds to the solution of
the equation $\
\partial \mathbf{J}^{ROMS}/\partial \delta z = 0$, and is given by:
\begin{equation}\label{primal}
\delta z^a = (\mathbf{B}^{-1} + G^T \mathbf{R}^{-1}G)^{-1}G^T \mathbf{R}^{-1}\mathbf{d}
\end{equation}
or, equivalently
\begin{equation}\label{dual}
\delta z^a =\mathbf{B}G^T (G\mathbf{B}G^T + \mathbf{R})^{-1}\mathbf{d}.
\end{equation}
Equation (\ref{primal}) is
referred to as the dual form (RBL4DVAR), while (\ref{dual}) is referred to as the primal form (IS4DVAR). In particular, we define 
\begin{equation}\label{kalman_gain}
    \mathbf{K}=\mathbf{B}G^T (G\mathbf{B}G^T + \mathbf{R})^{-1}
\end{equation}
as Kalman gain matrix.\\

\noindent TLROMS computes
\begin{equation}\label{TLROMS}
    \delta x^{ROMS}(t_l) \simeq M_{l-1,l}u(t_{l-1})
\end{equation}
where $\delta x^{ROMS}(t_l)=x^{ROMS}(t_l)-x^b(t_l)$, $\delta f(t_l)=f(t_l)-f^b(t_l)$, $\delta b(t_l)=b(t_l)-b^b(t_l)$, and $x^b(t_l)$, $f^b(t_)$, $b^b(t_l)$ are the background of the circulation, surface forcing and open boundary conditions respectively, and 
\begin{displaymath}
u(t_{l-1}) = ((\delta x^{ROMS})^T(t_{l-1}),\delta f^T(t_l),\delta b^T(t_l))^T.
\end{displaymath}
Equation (\ref{TLROMS}) is obtained from first-order Taylor expansion of NLROMS in (\ref{NLROMS}). \\ 

\noindent ADROMS computes 
\begin{equation}\label{ADROMS}
    u^*(t_{l-1}) = M_{l-1,l}^Tp(t_l)
\end{equation}
where $u^*(t_{l-1})=(p^T(t_{l-1}),\delta f^T(t_l),\delta b^{*T}(t_l))^T$ where $p$ is the adjoint state-vector, $\delta f^T$ and $\delta b^{*T}$ are the adjoint of the surface forcing and the open boundary condition increments.

\section{DD-4DVarDA in  ROMS model}\label{sec_DD}
Domain Decomposition (DD) method proposed in \cite{DD-4D} is made up of decomposition of the domain  $\Omega \times \Delta$ into subdomains where $\Omega$ is the 3D spatial domain and $\Delta$ is the time interval, solution of reduced forecast model and minimization of local 4D-Var functionals (we call it DD-4DVarDA method). \\
Relying on the existing software implementation, in the next we describe  main components of DD-4DVarDA method, highlighting the topics that we will address both on the mathematical problem underlying ROMS and the code implementation (see steps 1-4). We focus on IS4DVAR formulation described in Section \ref{ROMS}.
\subsection{Decomposition in space and time of the ocean model} 
\begin{figure}[!ht]
  \includegraphics[width=1.2\textwidth]{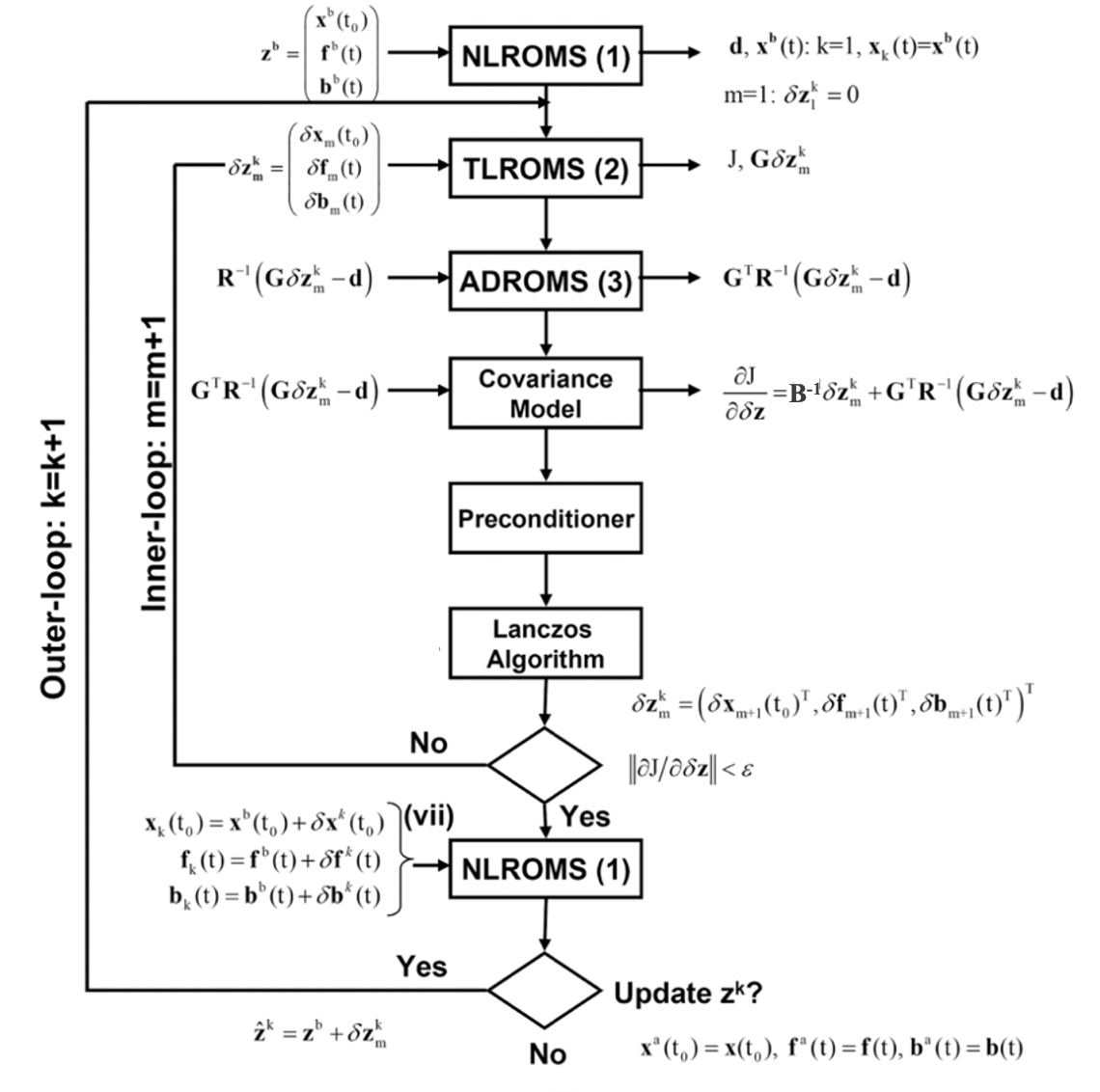}
  \caption{A flow chart illustrating IS4D-Var algorithm.}
\label{Fig_flow_chart}
\end{figure}

Decomposition of spatial domain $\Omega$.\\
We will consider a 2D decomposition of $\Omega\subseteq \mathbb{R}^3$ in x- and y-direction and denote 
 $\Omega_{xy}$ the spatial domain to decompose.\\

\begin{figure}[!ht]

 \centering
  \includegraphics[width=1.2\textwidth]{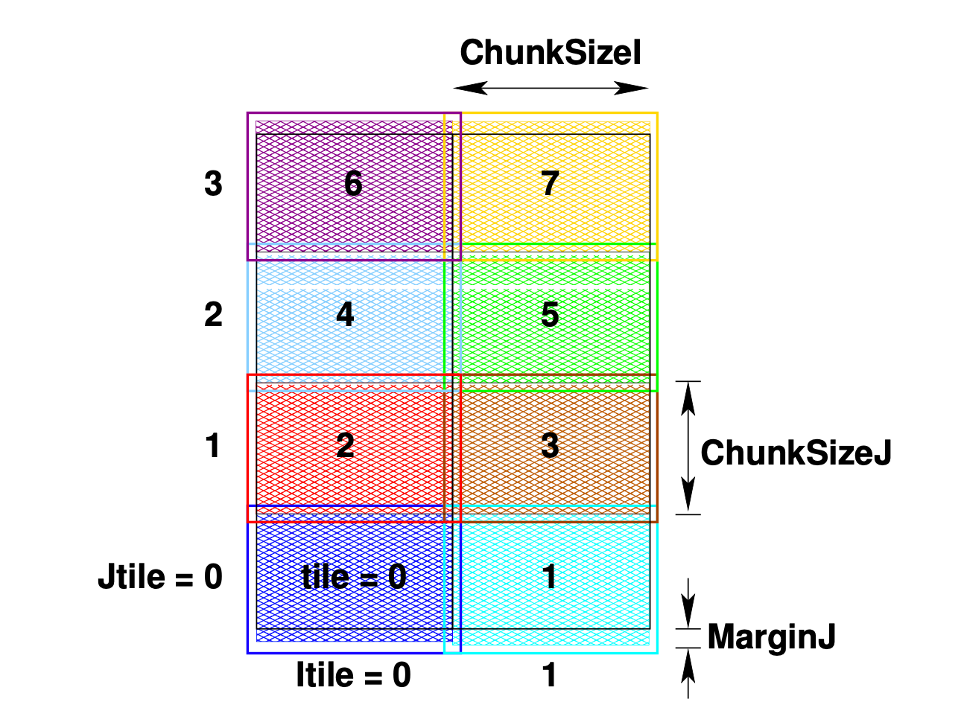}
   \caption{A tiled grid in xy-plane with some internal ROMS parameters.}
\label{Fig_tiles}
\end{figure}

 ROMS uses a parallelization approach that partitions domain $\Omega_{xy}$ into tiles (see Figure \ref{Fig_tiles}) \cite{myROMS}

\begin{equation}\label{spatial_dd}
    \Omega_{xy}=\bigcup_{i=0}^{N_{sub}-1}tile_{i}
    \end{equation}
    where $N_{sub}=NtileI \times NtileJ $; $NtileI$ and $NtileJ$ are the number of tiles set in the input file in x- and y-direction, respectively.\\ 
We denote by $HI$ and $HJ$ the overlapping tiles regions (i.e ghost or halo area in ROMS\footnote{In ROMS, the halo area would be two grids points wide unless the MPDATA advection scheme is used, in which case it needs three.}) in x- and y-direction, respectively.\\
\begin{tcolorbox}\textbf{{Step 0: DD of $\Omega$ in ROMS.\\}}
In our study we will assume the decomposition available in ROMS, as given in (\ref{spatial_dd}).
\end{tcolorbox}

\noindent Decomposition of time interval $\Delta$.\\
\noindent ROMS does not yet  implement  decomposition in time direction.\\
\begin{tcolorbox}\textbf{{Step 1: DD of $\Delta$ in ROMS.\\}}
     In our study we aim  introducing a decomposition of time interval $\Delta$ into $N_t$ intervals:
\begin{equation}\label{dd_time}
\Delta=\bigcup_{k=1}^{N_{t}}\Delta_{k}:=\bigcup_{k=1}^{N_{t}}[t_{\bar{s}_{k-1}},t_{\bar{s}_{k-1}+N_{k}}],
\end{equation}
where $N_{k}=|D(\Delta_{k})|$ are respectively the number of subdomains of $[0,T]$ and of time $t_l\in \Delta_{k}$ such that $\sum_{k=1}^{N_{t}}N_{k}-(N_{t}-1)=N$, $\bar{s}_{k-1}:=\sum_{j=1}^{k-1}N_{j}-(k-1)$ and $\bar{s}_{0}:=0$.
\end{tcolorbox}

\noindent Ocean model reduction.
ROMS allows each tile (or subdomain, see Figure \ref{Fig_tiles}) to compute local solutions of TLROMS and ADROMS.\\ For $i=0,1,\ldots, N_{sub}-1$ and $k=1,\ldots, N_t$ local\footnote{Let $x\in \mathbb{R}^{N_{p}}$ and $y\in \mathbb{R}^{N_{p}\times N}$ be vectors, for simplicity of notations, we refer to $x_{i}$ as a restriction of $x$ to $\Omega_{i}$, i.e. $x_{i}\equiv x/\Omega_{i}$ and $x_{i,k}\equiv x/(\Omega_{i}\times \Delta_k)$, similarly for matrix $A\in \mathbb{R}^{N_{p} \times N_{p}}$, i.e. $A_{i}\equiv A/\Omega_{i}$ and $A_{i,k}\equiv A/(\Omega_{i}\times \Delta_k)$, according the description in \cite{articolo2}.} TLROMS on local domain $tile_i$ computes
\begin{equation}\label{TLROMS_loc}
\delta x_i^{ROMS}(t_l) \simeq M_{i,(l-1,l)} u_i(t_{l-1})
\end{equation}
and local ADROMS on local domain $tile_i$ computes
\begin{equation}\label{ADROMS_loc}
    u_i^*(t_{l-1}) = M_{i,(l-1,l)}^T p_i(t_l),
\end{equation}
where $x_i$, $u_i$, $u_i^*$, $M_i$, $p_i$ are the restriction on $tile_i$ of variables $x$, $u$, $u^*$, $M$ and $p$ and $M_{i,(l-1,l)}$ is discrete model from $t_{l-1}$ to $t_l$.  \\
DD-4DVar method introduces  model reduction by using the background $x^b$ as local initial values. For $n=0,1,\ldots,\bar{n}$ (outer loop of DD--4DVAR method \cite{DD-4D}) do:
for $k=1,\ldots,N_{t}$, posed $x_{i,k}^{0}\equiv x_{i,k}^{b}$ $\forall i=0,1,\ldots,N_{sub}-1$, we let $x_{i,k}^{{M}_{i,k}}$ be the solution of the local model

\begin{equation}\label{co}
(P_{i,k}^{{M}_{i,k},n})_{i=0,1\ldots,N_{sub}-1,r=1,\ldots,N_{t}}\ : \quad 
\left\{\begin{array}{ll}
x_{i,k}^{\mathcal{M}_{i,k},n}=M_{i,k} x_{i,k-1}^n+b_{i,k} , 
\\
x_{i,k-1}^n=x_{i,k}^{\mathcal{M}_{i,k},n},
\\
x_{i,k-1}^n/HI=x_{i_I,k-1}^n/HI, \ \ (\ref{co}.1)\\
x_{i,k-1}^n/HJ=x_{i_J,k-1}^n/HJ, \ (\ref{co}.2)
\end{array}\right.
\end{equation}
where $i_I=0,\ldots,n_I-1$, $i_J=0,\ldots,n_J-1$, $n_I$ and $n_J$ are respectively numbers of adjacent tiles in x- and y-direction, $b_{i,r}^{k}$ and $M_i^{r}$ are respectively  background on $tile_i \times \Delta_{r}$, the vector accounting 
boundary conditions of $tile_{i}$ and the restriction in $tile_i$ of the matrix in (\ref{NLROMS}) that is
\begin{equation}\label{matriceM}
M_k\equiv M_{\bar{s}_{r-1},\bar{s}_{r}}:= M_{\bar{s}_{r-1},\bar{s}_{r-1}+1}\cdots M_{\bar{s}_{r}-1,\bar{s}_{r}}.
\end{equation}
In the following, we neglect the dependency on outer loop iteration $n$ of DD--4DVAR method.
We underline that local TLROMS and ADROMS in (\ref{TLROMS}) and (\ref{ADROMS}) are obtained by using MPI exchange for boundary conditions, regardless of local solution on overlap area, namely they do not consider overlapping tiles conditions in (\ref{co}.1) and (\ref{co}.2). 
Consequently, we need to modify local TLROMS and ADROMS in (\ref{TLROMS_loc}) and (\ref{ADROMS_loc}) taking into account of (\ref{co}.1) and (\ref{co}.1) conditions. More precisely,

\begin{tcolorbox}\textbf{{Step 2: TLROMS.}\\}
for $k=1,\ldots,N_{t}$, $\forall i=0,1,\ldots,N_{sub}-1$, local TLROMS on $tile_i \times \Delta_k$ will be modified such that
\begin{equation}\label{TLROMS_MOD}
\delta x_{i,k} \simeq M_{i,k} u_{i,k-1}+ \theta_{I}(u_{i,k-1})+ \theta_{J}(u_{i,k-1})
\end{equation}
where

\begin{equation}\label{vector_theta1}
\begin{array}{cc}
 \theta_{J}(u_{i,k-1}):=&\sum_{i_J=1}^{n_{J}} \gamma_{i_J}(M_{i,k}/HJ\cdot  u_{i,k-1}/HJ -\\ &M_k/HJ \cdot u_{i,k-1}/HJ) 
 \end{array}
\end{equation}

\normalsize
and

\begin{equation}\label{vector_theta2}
\begin{array}{cc}
 \theta_{I}(u_{i,r-1}):=&\sum_{i_I=1}^{n_{I}} \gamma_{i_I}(M_i^r/HI\cdot  u_{i,r-1}/HI -\\ &M_i^r/HI \cdot u_{i,r-1}/HI) 
  \end{array}
\end{equation}
are overlapping vectors in x- and y-direction, respectively; where parameters $\gamma_{i_I}$, $\gamma_{i_J}$ denote weights.
\end{tcolorbox}

Similarly, we need to modify local ADROMS in (\ref{ADROMS_loc}). More precisely, 
\begin{tcolorbox}\textbf{{Step 3: ADROMS.}\\}
for $k=1,\ldots,N_{t}$, $\forall i=0,1,\ldots,N_{sub}-1$, local ADROMS in (\ref{ADROMS_loc}) on local domain $tile_i \times \Delta_k$ will be modified such that 
\begin{equation}\label{ADROMS_mod}
u_{i,k-1}= (M_{i,r})^T \cdot p_{i,k}+\theta_{J}(p_{i,k-1})+\theta_{I}(p_{i,k-1})
\end{equation}
where $\theta_{J}$ and $\theta_{I}$ are defined in (\ref{vector_theta1}) and (\ref{vector_theta2}).

\end{tcolorbox}
\begin{figure}[!ht]
  
 \centering
  \includegraphics[width=1.2\textwidth]{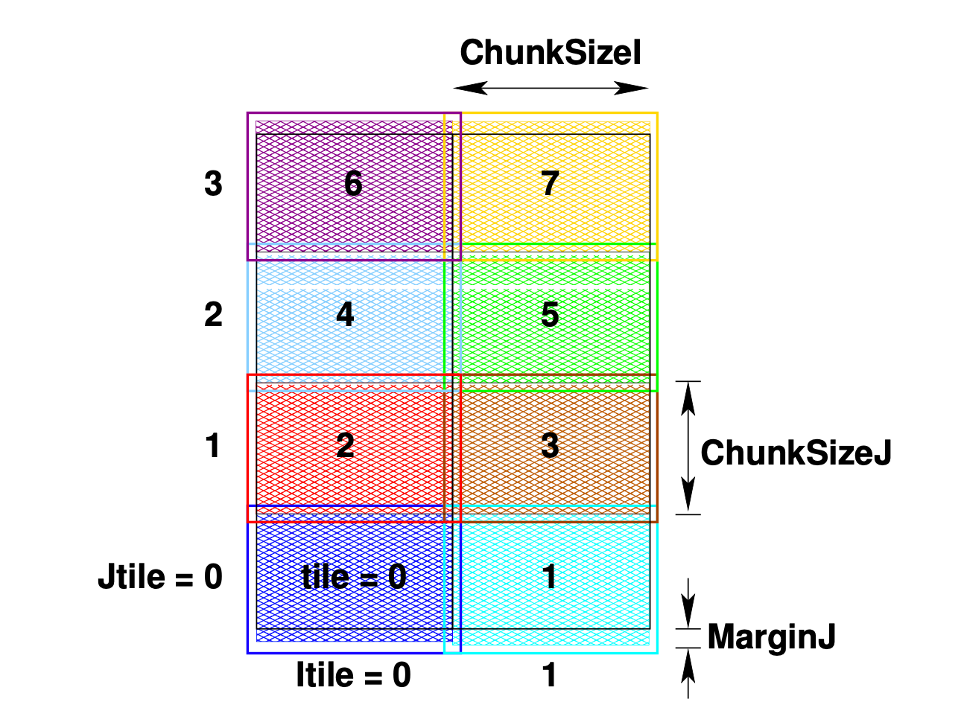}
  \caption{A region within the domain}
\label{Fig_region}
\end{figure}

\noindent  4D-VAR DA Operator Reduction.\\
\noindent Operator reduction involves the inner loop in Figure \ref{Fig_flow_chart}.\\
ROMS \cite{myROMS} computes and minimizes the operator 
 \begin{equation}\label{J_roms}
 \mathbf{J}_{tile_i}^{ROMS}:=  \mathbf{J}^{ROMS}/tile_i
  \end{equation}
 where $ \mathbf{J}^{ROMS}$ is defined in (\ref{funzionale}) and $ \mathbf{J}^{ROMS}/(tile_i)$ is the IS4D-Var functional in each tile $tile_i$ with MPI exchange of boundary conditions, $\forall i= 0,1,...,N_{sub}-1$.\\

Local 4D–VAR DA functional in \cite{DD-4D}  is defined as follows
\begin{equation}\label{local_fun}
 \mathbf{J}_{tile_i \times \Delta_r}^{DD-4DVar}:=  \mathbf{J}_{i,k}(x_{i,k})=\mathbf{J}(x_{i,k})/(tile_{i}\times \Delta_k)+\mathcal{O}_{IJ}(x_{i,k})
 \end{equation}
where
\begin{equation}\label{funzionale_4D}
 \mathbf{J}(x)=\alpha \|x-x^{b}\|_{\textbf{B}^{-1}}^{2}+\|Gx-y\|_{\textbf{R}^{-1}}^{2},
\end{equation}
is the DD-4DVar functional in \cite{DD-4D},
\begin{equation}\label{J_ris0}
\mathcal{O}_{IJ}(x_{i,k})=\sum_{i_I=1}^{n_{I}}\beta_{i_I,k}\cdot \|x_{i,k}/HI-x_{i_I,k}/HI\|_{\textbf{B}_{i_I}^{-1}}^{2}+\sum_{i_J=1}^{n_{J}} \beta_{i_J,k}\cdot \|x_{i,k}/HJ-x_{i_J,k}/HJ\|_{\textbf{B}_{i_j}^{-1}}^{2}
\end{equation}
is the overlapping operator on overlapping tiles region $HI$ and $HJ$, and  
\begin{equation}\label{J_ris1}
 \mathbf{J}_{i,k}(x_{i,k})/(tile_{i}\times \Delta_k)=  \alpha_{i,k} \cdot \|x_{i,k}-x_{i,k}^{{M}_{i,k}}\|_{\textbf{B}_{i}^{-1}}+\|{G}_{i,r}x_{i,k}-y_{i,k}\|_{\textbf{R}_{i}^{-1}}^{2}
\end{equation}
is the restriction of $ \mathbf{J}$ on $tile_{i}\times \Delta_k$, where $x^{b}$ is background, $y$ is observations vector in $\Delta$, $y_{i,k}$ is observations vector in $\Delta_k$; $\mathbf{B}_i$, $\mathbf{B}_{i_{I}}$, $\mathbf{B}_{i_J}$ are respectively the restrictions of covariance matrix $\mathbf{B}$ to $tile_i$, $HI$ and $HJ$; $G_{i,r}$, $\mathbf{R}_i$ are  restriction of matrices $G$ and $\mathbf{R}$ to $tile_i$ in $\Delta_k$. Parameters $\alpha_{i,r}$,  $\beta_{i_I}$ and $\beta_{i_J}$  in (\ref{J_ris0}) denotes  regularization parameters. We let $\alpha_{i,k}=\beta_{i_I,k}=\beta_{i_J,k}=1$, $\forall i_I=1,\ldots,n_{I}$ and $\forall i_J=1,\ldots,n_{J}$.\\
Incremental formulation of local  4D–VAR  DA  functional  in (\ref{local_fun}) is $$ {J}_{i,k}(\delta z_{i,k})={J}(\delta z_{k})/(tile_{i}\times \Delta_k)+\mathcal{O}_{IJ}(\delta z_{i,k})$$ where $\delta z_{i,k}$ are the control variable increments in $tile_i\times\Delta_k$.
\\ 
\begin{tcolorbox}\textbf{{Step 4: IS4D-Var.}\\}
From (\ref{funzionale_4D}) local IS4D-Var function in (\ref{J_roms}) can be written \begin{equation}\label{J_mod}
   \mathbf{J}_{loc}^{ROMS}=\mathbf{J}_{tile_i \times \Delta}^{DD-4DVar}-\mathcal{O}_{IJ}
\end{equation}

Consequently, we need to add overlapping operator in order to enforce the matching of local solutions on the overlapping tiles in each time interval. 
\end{tcolorbox}

\begin{figure}[!ht]
 \centering
 
  \includegraphics[width=1.\textwidth]{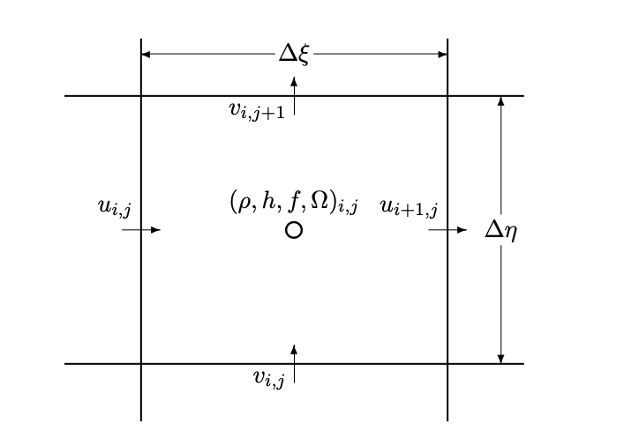}
  \caption{ Placement of variables on an Arakawa C grid. }
\label{Fig_grid}
\end{figure}

\begin{figure}[!ht]
 
 \centering
  \includegraphics[width=1.\textwidth]{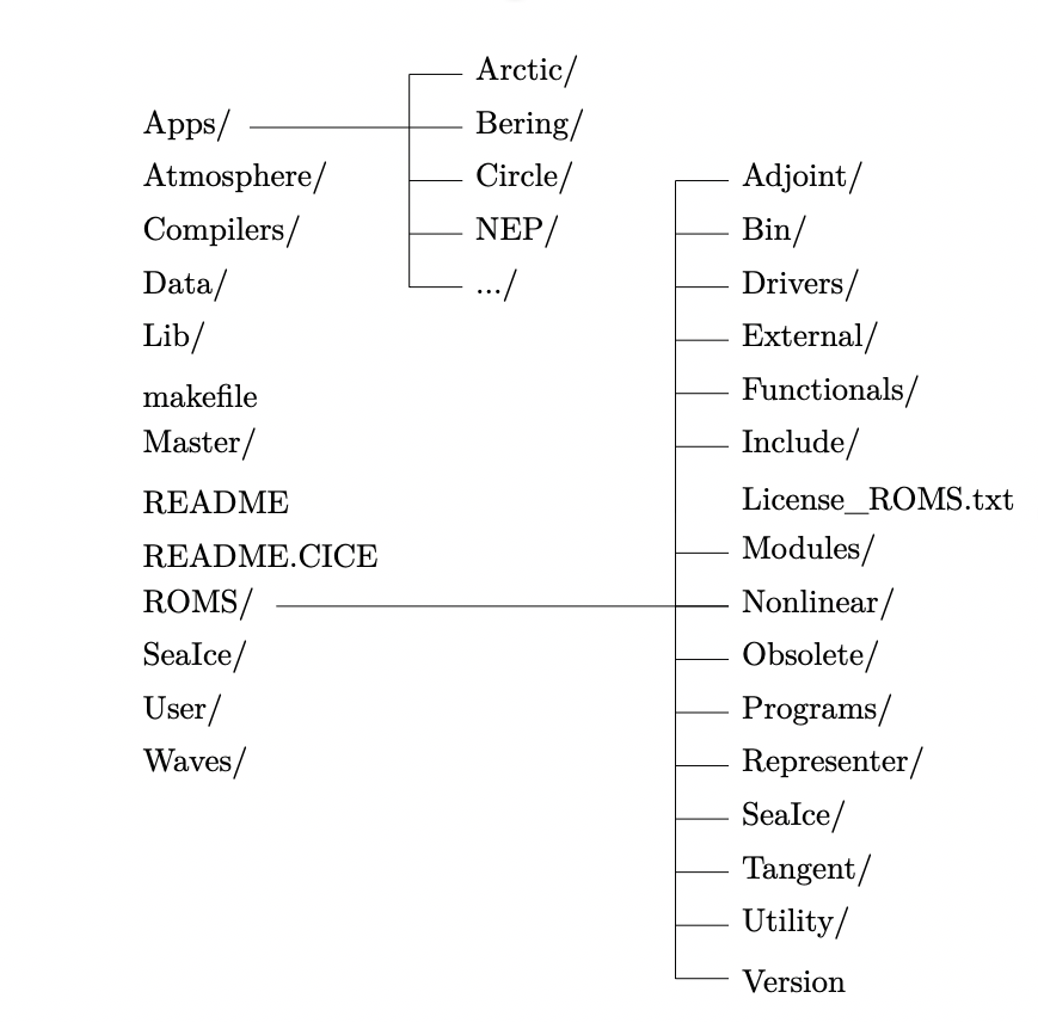}
 \caption{ ROMS directory structure. }
\label{Fig_directory}
\end{figure}

\subsection{DD-4DVarDA in ROMS code}\label{DD-4dvar_roms}
In Figure \ref{Fig_directory} the ROMS directory structure is shown. We focus on ROMS folder, in particular, on its folders: Tangent and Adjoint. 
\begin{enumerate}
\item ROMS.\\
We need to modify routines in ROMS folder implementing decomposition as in (\ref{dd_time}) (see step 1).
\\
\noindent 
Decomposition of time interval involves modification of initial conditions of nonlinear model, in TLROMS and ADROMS.\\
Routines involving initialization are (see Figure \ref{Fig_directory}):
\begin{enumerate}
    \item initial routine in Nonlinear folder: initializes all model variables (it is called in main3d).
    \begin{tcolorbox}\textbf{Note 1.1 (see step 1)\\}
Initial routine initializes all model variable before  calling main3d routine (in Nonlinear folder). Main3d routine is the main driver solving nonlinear ROMS model (background).  \\

DD in x- and y- directions is not applied for computing background, i.e. there are not MPI communications among processes.  \\This means that background is computed without using DD in space, consequently we not apply DD in time.\\ 
Moreover, some values of $u,v$ (components velocity) are not set to zero at the end of time step, because its values on some grid points are necessary for the next time step.

\end{tcolorbox}
    \item tl\_initial routine in Tangent folder: initializes all tangent model variables (it is called in i4dvar).
    \begin{tcolorbox}\textbf{{Note 1.2 (see step 1).\\}}
We consider tl\_initial routine. Parts involving initializations are:
\begin{itemize}
    \item line 123:  Initializes time stepping indices and counter.
    \item line 162:  initialization time.

\end{itemize}
\end{tcolorbox}

    \item ad\_initial routine in Adjoint folder: initializes all adjoint model variables(it is called in i4dvar).
    \begin{tcolorbox}\textbf{{Note 1.3 (see step 1).\\}}
We consider ad\_initial routine. Parts involving initializations are:
\begin{itemize}
    \item line 113:  Initializes time stepping indices and counter.
    \item line 152:  initialization time.
\end{itemize}
\end{tcolorbox}
\end{enumerate}
Main actions to apply DD in time in ROMS are described below. 
\begin{tcolorbox}\textbf{{Action 1.1 (see step 1).\\}}
\begin{enumerate}

\item Modify initials routines by adding MPI communications for initial conditions to tangent and adjoint routines in each time interval. \\ 
\item Decompose time interval (RunInterval variable in routines).
One possible way forward:
    \begin{itemize}
\item We need to use OMP threadprivate directive for replicating variables related to time interval such that each thread has its own copy. (see mod\_parallel routine in Module folder at line 51 related to DD in space).\\
Allocate\_routine in mod\_parallel allocates tiles; equally we can allocate local time interval and related variables. \\
\item We can add allocation and  OMP threadprivate directive of local time intervals in mod\_parallel routine. 
We can define first time interval (first\_time\_interval) and last time interval (last\_time\_interval) and add a \textit{for loop}, after \textit{for loop} involving tiles, started from first time interval up to last time interval adding a time step (dt) defined in driver. \\
\begin{itemize}
    \item tl\_main3d: at line 277 stars \textit{for loop} involves tiles.
    \item ad\_main3d: at line 629 stars \textit{for loop} involves tiles.
\end{itemize}
\item Identify tangent and adjoint variables for MPI communications in time. 

\end{itemize}

\item Introduce MPI communications in time.
\begin{itemize}
\item We need to split ROMS MPI communicator (OCN\_COMM\_WORLD) for obtaining MPI communicators needed to communications among processes related to same spatial sub domain but different time intervals.\\
\noindent By splitting ROMS MPI communicator we obtained a new MPI communicator namely one single communicator for the current process.

 \end{itemize}
\end{enumerate}

\end{tcolorbox}

\begin{tcolorbox}\textbf{{Action 1.2 (see step 1).\\}}
Step 1 involves:
\begin{itemize}
\item  tl\_main3d routine: at line 142 starts \textit{while loop} on time interval (RunInterval) by increasing the step time (my\_steptime). Moreover, at lines 278 and 280 it calls tl\_set\_massflux and tl\_rho\_eos, i.e. the routines we need to modify (see Action 2.1,2,3,4). Consequently, we probably need to introduce in tl\_main3d MPI communications in time after the tl\_set\_massflux and tl\_rho\_eos routines.

\item ad\_main3d routine: at line 177 starts \textit{while loop} on time interval (RunInterval) by increasing the step time (my\_steptime). Moreover, at lines 632 and 634 it calls ad\_rho\_eos and ad\_set\_massflux, i.e. the routines we need to modify (see Action 3.1 and 3.2). Consequently, we probably need to introduce in ad\_main3d MPI communications in time after the ad\_set\_massflux and ad\_rho\_eos routines.
\end{itemize}

\end{tcolorbox}

    \item TLROMS.\\
We need to modify TLROMS as in (\ref{TLROMS_MOD}) (see step 2), namely we need to compute the overlapping vector in (\ref{vector_theta1}) and (\ref{vector_theta2}) and add them to tangent variables. \\
\begin{tcolorbox}\textbf{{Action 2.1 (see step 2).}\\}
We could consider another inner loop (inside the loop over $m$ in Figure \ref{Fig_flow_chart}) over index $n$ and  initial approximation of solution on local tiles at $n=0$.
For each iteration we need local solution on tile adjacent to each tile, this means using MPI exchange of information between adjacent tiles (in two direction see Figure \ref{Fig_grid}). 

\end{tcolorbox}
We note that primitive equations of motion \cite{ROMS_TLROM} are written in flux form transformed using orthogonal curvilinear coordinates ($\xi,\eta$) (see Figure \ref{Fig_grid}).\\
We consider tl\_main3d routine. In tangent folder (see Figure \ref{Fig_directory}) this routine is the main driver of TLROMS configurated as a full 3D baroclinic ocean model.\\ tl\_main3d routine calls the following subrountines.
\begin{enumerate}
\item 

tl\_set\_massflux calls tl\_set\_massflux\_tile. \\
tl\_set\_massflux\_tile: computes “in situ” tangent linear horizontal mass flux. 
\begin{tcolorbox}\textbf{{Action 2.2 (see step 2).\\}}
Taking into account (\ref{vector_theta1}) and (\ref{vector_theta2}) we need to modify the code starting from line 155.

\end{tcolorbox}

\item tl\_rho\_eos calls tl\_rho\_eos\_tile. \\
tl\_rho\_eos\_tile: computes “in situ” the density and other quantities (temperature, salinity,\ldots).\\
\begin{tcolorbox}\textbf{{Action 2.3 (see step 2).\\}}
Parts to be modified:
\begin{enumerate}
    \item Line 505: computes "in situ" density anomaly;
\item Line 591: computes "in situ" Brunt-Vaisala frequency;
\item Line 1158: computes "in situ" Brunt-Vaisala frequency;

\end{enumerate}
\end{tcolorbox}
\item
tl\_omega: computes vertical velocity (no modifications because DD is only horizontal, there are not overlap region along vertical direction).
\end{enumerate}

\item ADROMS.\\
Similarly to TLROMS, we need to modify ADROMS as in (\ref{ADROMS_mod}) (see step 3), namely we need to compute and add the overlapping vector in (\ref{vector_theta1}) and (\ref{vector_theta2}) to adjoint variables. \\
We consider ad\_main3d routine.
In Adjoint folder (see Figure \ref{Fig_directory}) this routine is the main driver of ADROMS configurated as a full 3D baroclinic ocean model. \\
ad\_main3d routine calls the following subrountines.
\begin{enumerate}
    \item ad\_rho\_eos: computes “in situ” density and other associated quantities.\\
    \begin{tcolorbox}\textbf{{Action 3.1 (see step 3).\\}}
Parts that should be modified:

\begin{enumerate}
    \item Lines 797 and 1758: compute "in situ" adjoint Brut-Vaisala frequency at horizontal points.
     \item Line 1070: computes "in situ" adjoint  density anomaly.
     \end{enumerate}
\end{tcolorbox}
\item	ad\_set\_mass\_flux: compute “in situ” adjoint horizontal mass fluxes.\\
 \begin{tcolorbox}\textbf{{Action 3.2 (see step 3).\\}}
Part that should be modified:
\begin{enumerate}
\item Line 201: computes "in situ" adjoint horizontal mass fluxes

\end{enumerate}
\end{tcolorbox}
\item	ad\_set\_avg: accumulates and computes output time-averaged adjoint fields. (probably no modifications are needed).
\end{enumerate}

\item IS4D-Var.\\
We need to modify routines in ROMS folder as in (\ref{J_mod}) (see step 4).\\
\begin{tcolorbox}\textbf{{Action 4.1 (see step 4).\\}}
We need to modify IS4DVAR cost  function in (\ref{funzionale}) to take in account overlap region i.e. halo region. 
\end{tcolorbox}

\end{enumerate}


\section{ROMS test cases}
\noindent We have installed ROMS on the high-performance hybrid computing architecture of the Sistema Cooperativo Per
Elaborazioni scientifiche multidiscipliari data center, located in the University of Naples Federico II. More precisely, the HPC architecture is made of 8 nodes, consisting of distributed memory DELL M600 blades connected by a 10 Gigabit Ethernet technology. Each blade consists of 2 Intel Xeon@2.33 GHz quadcore processors sharing 16 GB RAM memory for a total of 8 cores/blade and of 64 cores, in total.\\
We consider test cases referred to IS4D-Var and  RBL4DVAR  configured for the U.S. west coast and the California Current System (CCS). This configuration, referred to as WC13, has 30 km horizontal resolution, and 30 levels in the vertical and data assimilated every 7 days during the period July 2002 – Dec. 2004.

\begin{figure}
{\includegraphics[width=.5\textwidth]{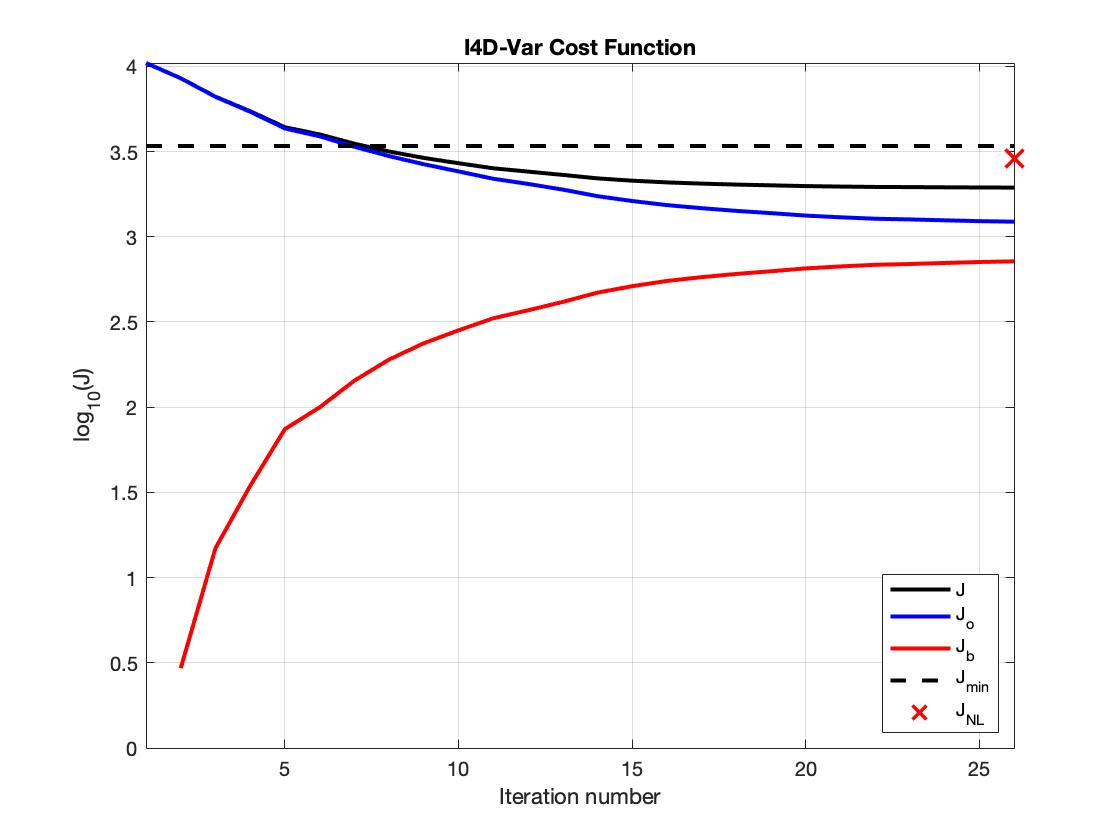}}
\quad
{\includegraphics[width=.5\textwidth]{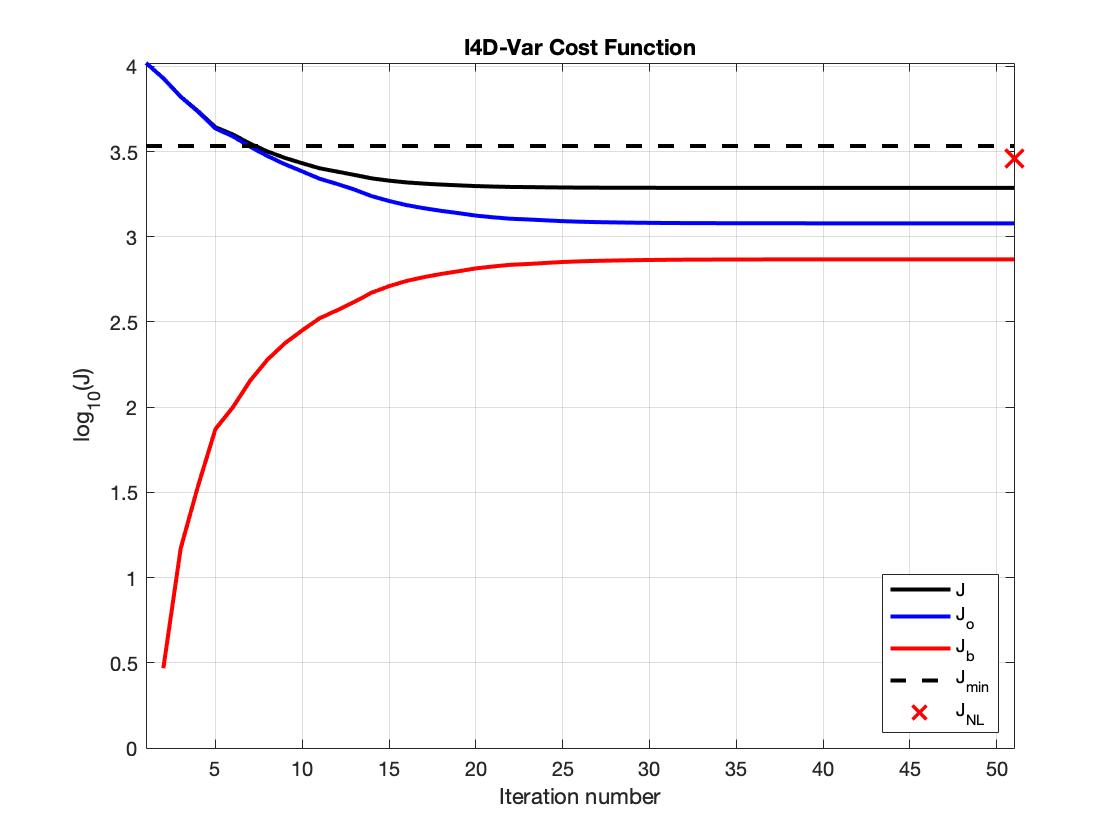}}\quad
{\includegraphics[width=.5\textwidth]{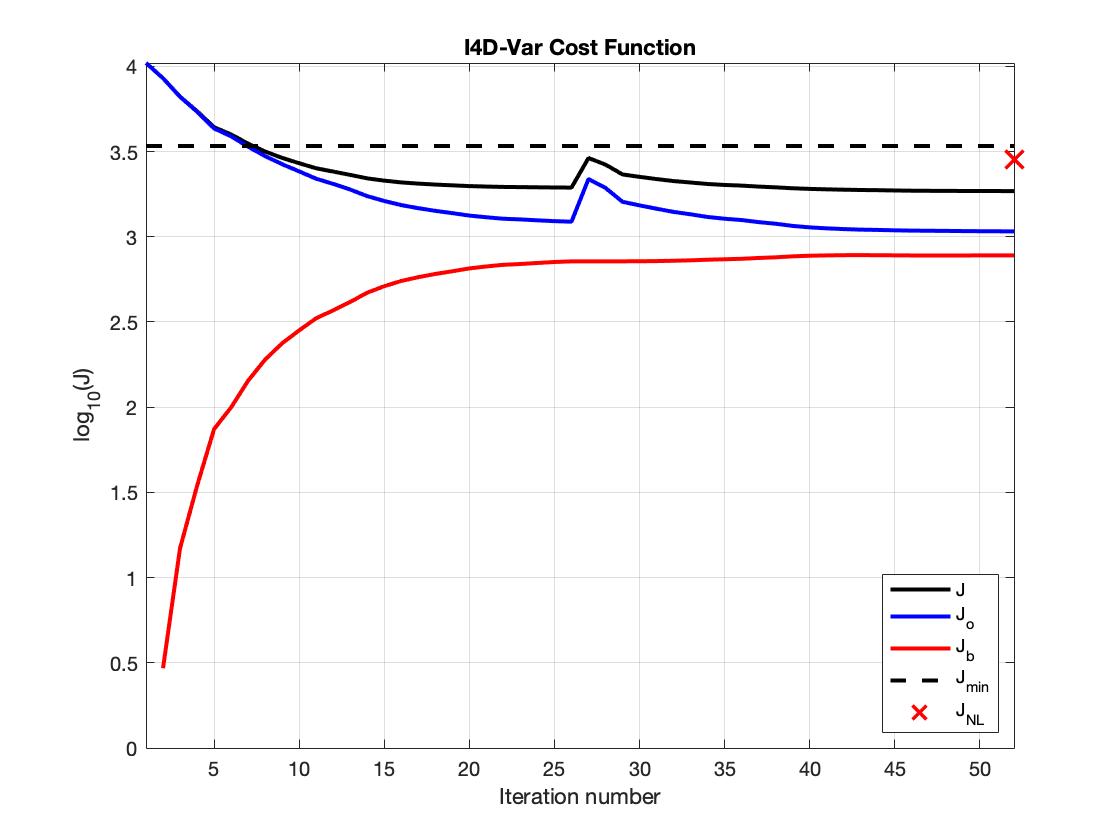}}\quad
{\includegraphics[width=.5\textwidth]{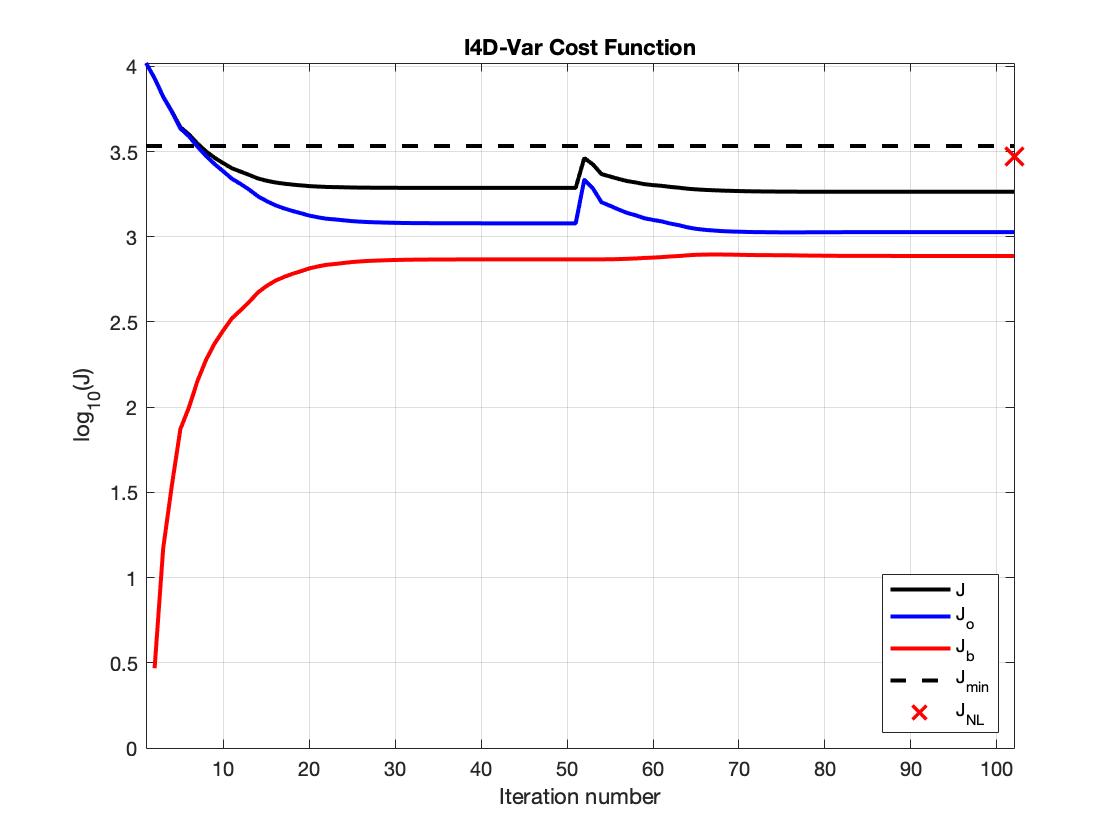}}\\
{\caption{IS4D-Var Cost Function: case1, case2, case3 and case 4, respectively, using roms\_wc13\_2hours.in input file.}

\label{grafico_roms2}}
\end{figure}

\begin{figure}
{\includegraphics[width=.5\textwidth]{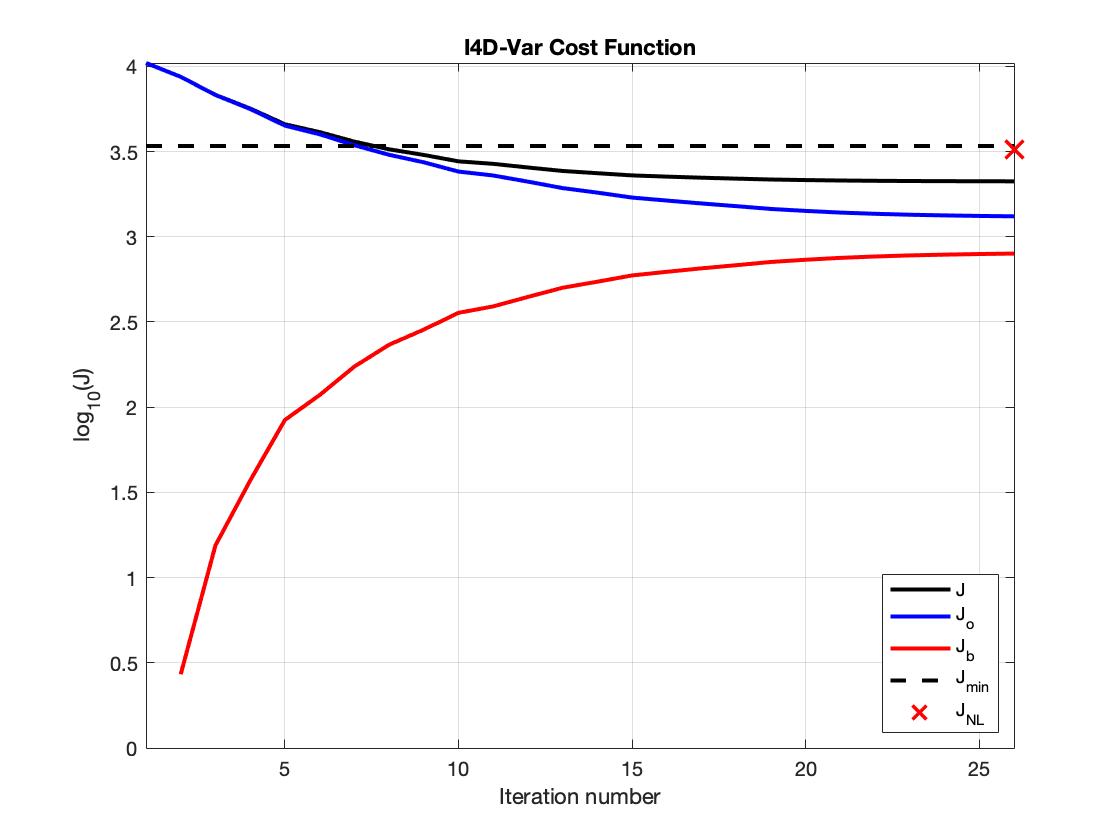}}
\quad
{\includegraphics[width=.5\textwidth]{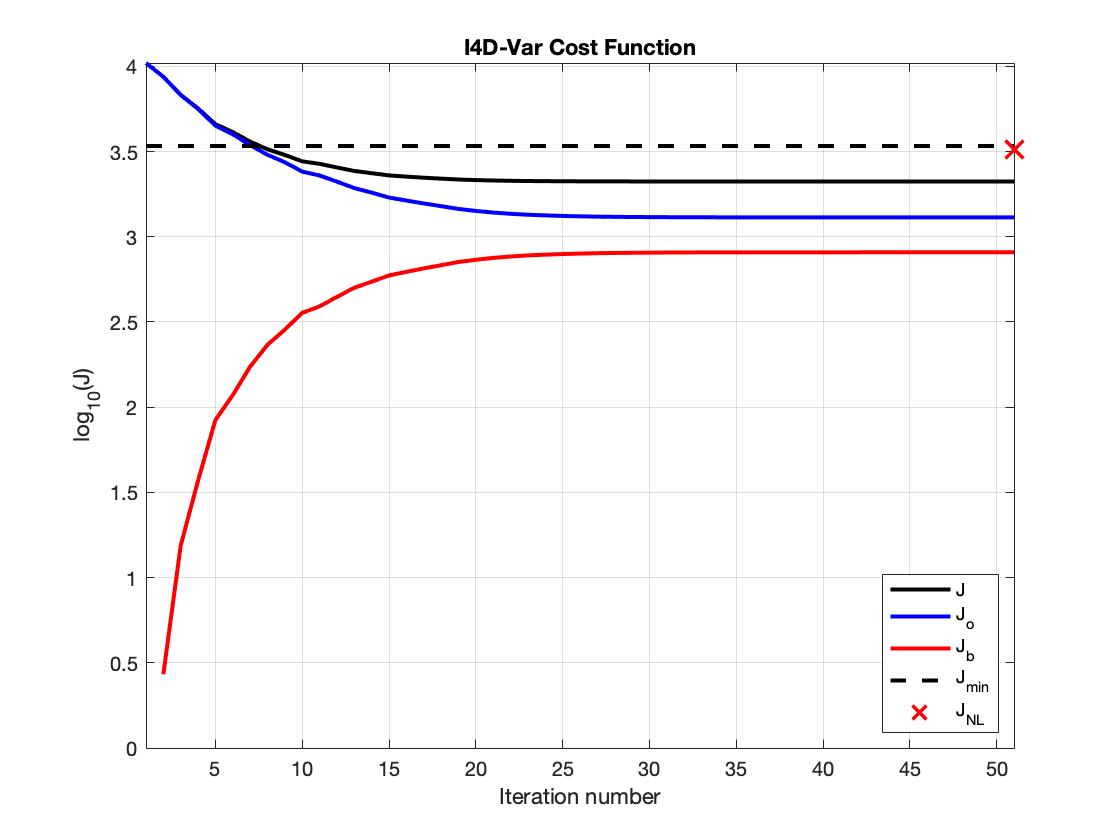}}\quad
{\includegraphics[width=.5\textwidth]{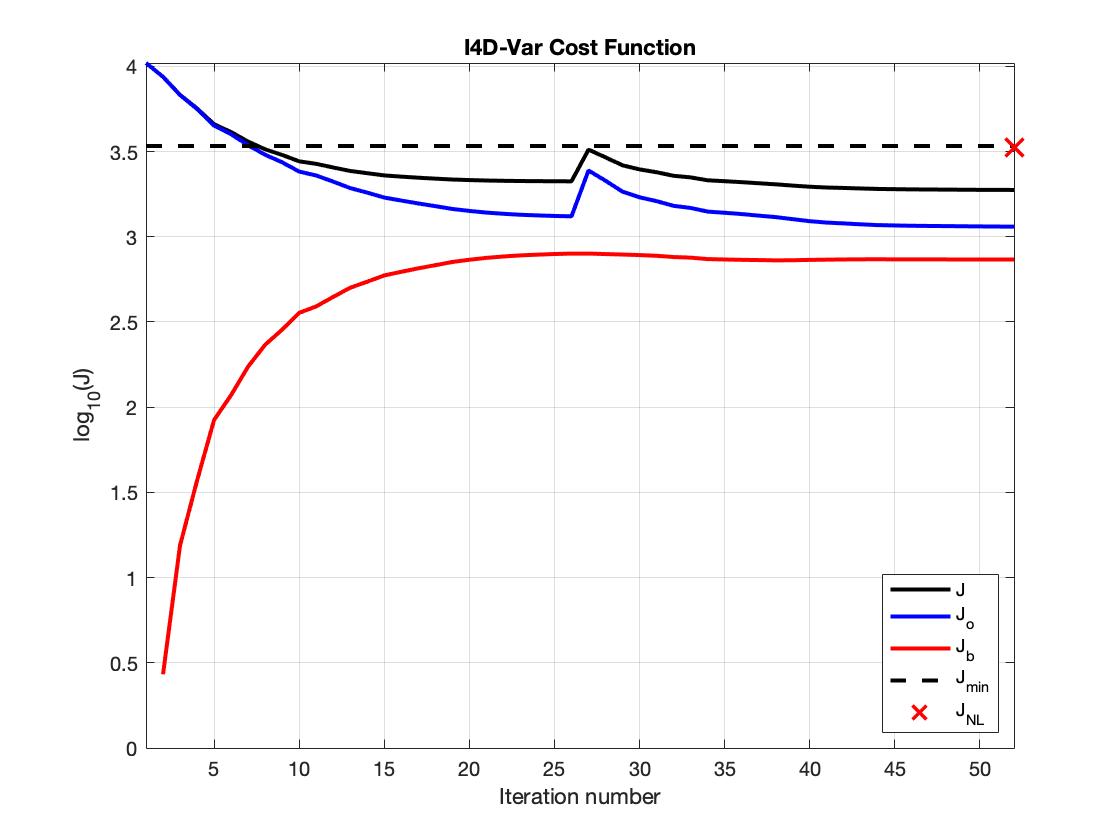}}\quad
{\includegraphics[width=.5\textwidth]{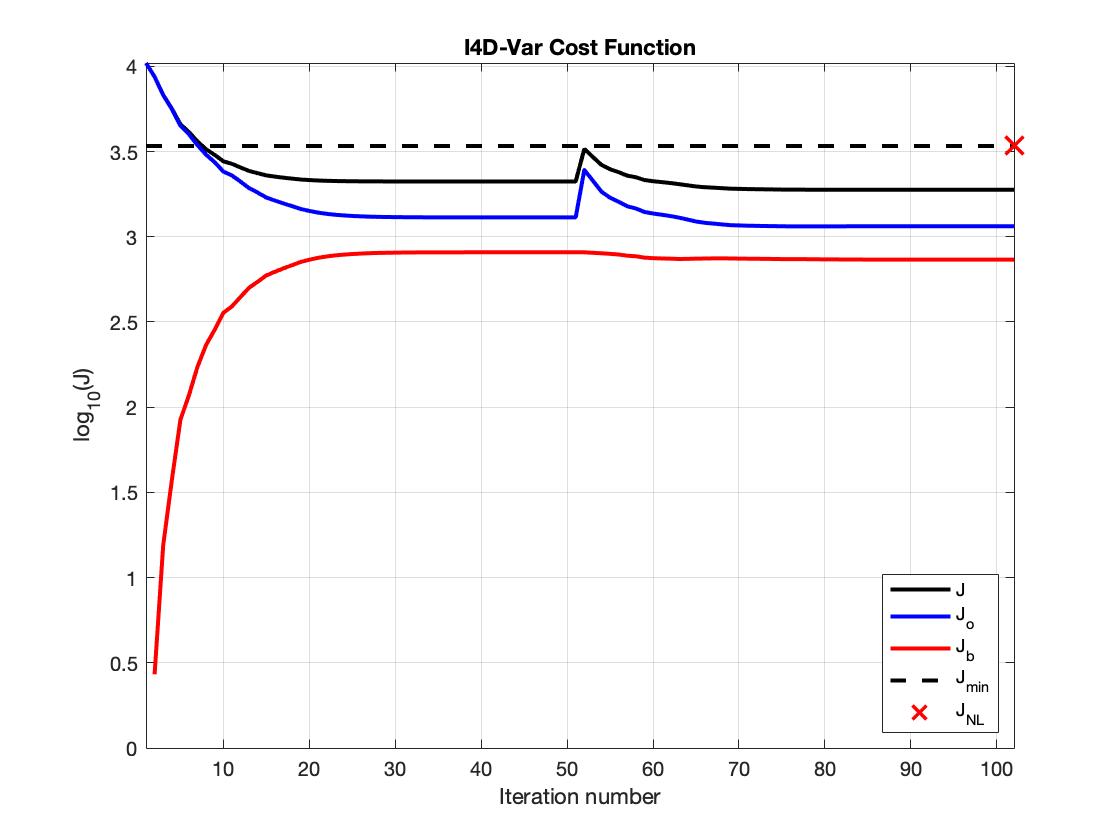}}\\
{\caption{IS4D-Var Cost Function: case1, case2, case3 and case 4, respectively, using roms\_wc13\_daily.in input file.}

\label{grafico_roms1}}
\end{figure}

\begin{itemize}
    \item IS4DVAR test cases.\\

\noindent To run this application I need to take the following steps.
\begin{itemize}
    \item I customize build\_roms.csh script in IS4DVAR folder and define the path to the directories where all project's files are kept.
    \begin{itemize}
        \item   I set USE\_MY\_LIBS to yes, then I modify source code root file Complilers/my\_build\_paths.sh and edit the appropriate paths for the desired compiler. 
    \end{itemize}
 \item I customize the ROMS input file (roms\_wc13\_2hours.in/roms\_wc13\_daily.in) and specify
      the appropriate values of tile numbers in the I-direction and J-direction i.e. NtileI=2 and NtileJ=4 (see Figure \ref{Fig_tiles}).
\item I modify ARmake.inc in Lib folder to create the libraries for serial and parallel ARPACK.
\item I customize Linux-ifort.mk (Fortran compiler is ifort) and define library locations.
\item I create PBS scripts to run the test cases.

\end{itemize}
Figures \ref{grafico_roms1} and \ref{grafico_roms2} show results obtained applying DD in space by considering $NtileI\times NtileJ$ tiles and input files roms\_wc13\_2hours.in and roms\_wc13\_daily.in, respectively. \noindent
They show total cost function $J$ (black curve), observation cost function $Jo$ (blue curve), and background cost function $Jb$ (red curve) and the theoretical minimum value $J_{min}=n_{obs}/2$ (dashed black line) plotted on a $log_{10}$ scale. We observe  that the solution has more or less converged by 25 iterations.
\\We will refer to the following quantities: Nouter number of outer loops, Ninner inner loop, $T_{Casei}^{node}$ elapsed time for each node in seconds, $T_{max}$ and $T_{min}$ maximum and minimum elapsed time among nodes in seconds.
\\
We consider:
\begin{itemize}
    \item case 1: Nouter=1 and Ninner=25;
    \item case 2: Nouter=1 and Ninner=50;
    \item case 3: Nouter=2 and Ninner=25;
    \item case 4: Nouter=2 and Ninner=50;
\end{itemize}
In Tables \ref{tab_2hours}, \ref{tab_max_min_1} and \ref{tab_daily}, \ref{tab_max_min_2} we report elapsed time for each process considering roms\_wc13\_2hours and roms\_wc13\_daily.in input file, respectively. \\

\begin{table}
\normalsize
\centering
\begin{tabular}{rlllllllllll}
\hline
 $Node$ &  $T_{Case1}^{node}$ & $T_{Case2}^{node}$ & $T_{Case3}^{node}$ &$T_{Case4}^{node}$  \\
\hline
$0$ &$1.56\times 10^3$ &$3.24\times 10^3$& $3.124\times 10^3$&$6.42\times 10^3$\\
$1$ & $1.58\times 10^3$&$3.29\times 10^3$& $3.17\times 10^3$&$6.49\times 10^3$\\
$2$ &$1.58\times 10^3$&$3.29\times 10^3$& $3.17\times 10^3$&$6.50\times 10^3$\\
$3$ &$1.58\times 10^3$& $3.28\times 10^3$&$3.16\times 10^3$&$6.50\times 10^3$\\
$4$ & $1.58\times 10^3$&  $3.29\times 10^3$&$3.17\times 10^3$&$6.50\times 10^3$\\
$5$ & $1.58\times 10^3$&$3.29\times 10^3$&$3.16\times 10^3$&$6.50\times 10^3$\\
$6$ & $1.58\times 10^3$&$3.29\times 10^3$&$3.16\times 10^3$&$6.50\times 10^3$\\
$7$ &$1.58\times 10^3$& $3.28\times 10^3$&$3.17\times 10^3$&$6.49\times 10^3$\\

\hline 
\end{tabular}
\caption{\footnotesize Elapsed time in seconds for each process related  to roms\_wc13\_2hours input file.}
\label{tab_2hours}
\end{table}

\begin{table}

\centering
\begin{tabular}{rlllllllllll}
\hline
& $T_{max}$ & $T_{min}$  \\
\hline
${Case1}$& $1.58\times 10^3$ &$1.56\times 10^3$\\
$Case2$&$3.29\times 10^3$ &$3.24\times 10^3$ \\
${Case3}$&$3.17\times 10^3$ &$3.12\times 10^3$\\
${Case4}$&$6.50\times 10^3$ &$6.42\times 10^3$\\

\hline 
\end{tabular}
\caption{Maximum and minimum elapsed time in seconds among nodes related to roms\_wc13\_2hours.in input file.}
\label{tab_max_min_1}
\end{table}

\begin{table}

\normalsize
\centering
\begin{tabular}{rlllllllllll}

\hline
 $Node$ &  $T_{Case1}^{node}$ & $T_{Case2}^{node}$ & $T_{Case3}^{node}$ &$T_{Case4}^{node}$  \\
\hline
$0$ & $1.41\times 10^3$ &$2.95\times 10^3$& $2.79\times 10^3$&$5.97\times 10^3$\\
$1$  & $1.41\times 10^3$ &$2.96\times 10^3$&$2.83\times 10^3$&$5.99\times 10^3$\\
$2$  & $ 1.41\times 10^3$& $2.97\times 10^3$&$2.82\times 10^3$&$5.99\times 10^3$ \\
$3$  &$1.41\times 10^3$&$2.97\times 10^3$& $2.82\times 10^3$ & $6.00\times 10^3$\\
$4$  & $1.41\times 10^3$ &$2.97\times 10^3$&$2.82\times 10^3$&$5.98\times 10^3$\\
$5$  &$1.41\times 10^3$& $2.96\times 10^3$&$2.82\times 10^3$&$5.99\times 10^3$\\
$6$  & $1.41\times 10^3$ & $2.96\times 10^3$&$2.82\times 10^3$&$6.00\times 10^3$\\
$7$  & $1.41\times 10^3$&$3.00\times 10^3$&$2.82\times 10^3$&$6.00\times 10^3$\\
\hline 
\end{tabular}
\caption{Elapsed time in seconds for each process releted to roms\_wc13\_daily.in input file.}
\label{tab_daily}
\end{table}

\begin{table}

\centering
\begin{tabular}{rlllllllllll}
\hline
& $T_{max}$ & $T_{min}$  \\
\hline
${Case1}$ &$1.41\times 10^3$ &$1.41\times 10^3$\\
$Case2$  &$3.00\times 10^3$ &$2.95\times 10^3$\\
${Case3}$ & $2.83\times 10^3$ &$2.79\times 10^3$\\
${Case4}$  & $5.97\times 10^3$ &$5.97\times 10^3$\\

\hline 
\end{tabular}
\caption{ Maximum and minimum elapsed time in seconds among nodes  related to roms\_wc13\_daily.in input file.}
\label{tab_max_min_2}
\end{table}

\item RBL4DVAR.\\

\noindent To run this application I need to take the following steps.
\begin{itemize}
    \item I customize build\_roms.csh scripts in RBL4DVAR folders and define the path to the directories where all project's files are kept.
 \item I customize the ROMS input file (roms\_wc13\_2hours.in/roms\_wc13\_daily.in) and specify
      the appropriate values of tile numbers in the I-direction and J-direction i.e. NtileI=2 and NtileJ=4 (see Figure \ref{Fig_tiles}).

\item I create a PBS script to run the test cases.

\end{itemize}

    \item IS4D-Var vs RBL4DVAR.
    \\

IS4D-Var and RBL4DVAR algorithms are based  
on a search for the best circulation estimate in the space spanned by the model control vector and in the dual space spanned by the observations, respectively. We note that ${n_{obs}}<<N_{p}$, where ${n_{obs}}$ is number of observations and $N_{p}$ is dimension of model control vector. 
Hence, the dimension of observation space is significantly smaller than the model control space, the dual formulation can reduce both memory usage and computational cost. 
\noindent Consequently, it appears that the dual formulation should be an easier problem to solve because of the considerably smaller dimension
of the space involved, but it has practical barriers to convergence. \\  \noindent Consequently, we consider three applications of dual formulation:
\begin{enumerate}

\item using the standard $R^{-1/2}$ preconditioning with a conjugate gradient (RBL4DVAR) method;
   \item using the $R^{-1/2}$ preconditioning and  minimum residual algorithm (MINRES);
   \item using a restricted B-preconditioned conjugate gradient (RPCG) approach.
\end{enumerate}

Figure \ref{grafico_roms_rbl4dvar} and \ref{grafico_roms_rbl4dvar_2} shows the convergence of minimization algorithms RBL4DVAR and RBL4DVAR, MINRES, and RPCG compared to primal formulation IS4DVAR by considering $Nouter=1$, $Ninner=50$ and $Nouter=1$, $Ninner=25$. We observe that the performance of RPCG is
superior to both MINRES and RBL4DVAR, in particular, PCG ensures that RPCG converges at same rate as IS4D-Var.

\noindent
\item Observation impact and  observation sensitivity.\\
The observations assimilated into the model are:
\begin{itemize}
 \item  %
 SST satellite: Sea Surface Temperature;
    \item 
    SSH satellite: Sea Surface Heights;
    \item  Argo floats:  hydrographic observations of temperature and salinity.
\end{itemize}   
We consider the time average transport
across 37N over the upper 500 m, denoted by $I_{37N}$, and given by
\begin{equation}\label{average_37N}
    I_{37N} (x) = \frac{1}{N} \sum_{i=1}^{N}h^Tx_i
\end{equation}
where $h$ is a vector with non-zero elements corresponding to the velocity grid points that
contribute to the transport normal to the 37N section shown in Figure \ref{domain_obs}, $N$ is the number of time steps during the assimilation interval, $x_i$ is the model state-vector at time $i\Delta t$, and $\Delta t$ is the model time step. We consider scalar functions of ocean state vector $I=I(x)$, prior $I^b=I(x^b)$ and posterior ocean state vector $I^a=I(x^a)$, where $x^b$ is background circulation estimate and $x^a$ is 4D-Var analysis. The circulation analysis increment $\delta x^a(t)$ at instant time $t$ in interval time  $[t_0, t_0 + 7]$ of each 4D-Var cycle is given
\begin{equation}\label{incremet_analysis}
   \delta x^{ROMS}(t)=x^a (t)-x^b(t),
\end{equation}
consequently, transport increment $\Delta I$ during each assimilation cycle can be expressed as:
\begin{equation}\label{increment}
   \Delta I=I^a -I^b= I(x^b +\delta x)-I^b \simeq  \delta x^T (\partial J/\partial x),
\end{equation}
using the tangent linear assumption in  (\ref{TLROMS}), (\ref{kalman_gain}) and (\ref{dual}), the circulation analysis increment is given by
\begin{equation}
\delta x^{ROMS}(t)=M(t,t_0 )\Tilde{\mathbf{K}} d,
\end{equation}
where $M(t,t_0)$ represents the perturbation tangent linear model for the time interval $[t_0,t]$, $\Tilde{\mathbf{K}}$ is an approximation of 
$\mathbf{K}$ defined in (\ref{kalman_gain}) and $\mathbf{d}$ is innovation vector. Then, increment defined in (\ref{increment}) become 
\begin{equation}\label{formulation_increment}
    \Delta I\simeq d^T\Tilde{K}^TM^T(\partial I/\partial x),
\end{equation}
where $\Tilde{K}$ represents adjoint of Kalman gain matrix. \\
Consequently, 37N time increment of averaged transport in (\ref{average_37N}) is: 
\begin{equation}
\Delta I_{37N}\simeq \frac{1}{N}d^T\Tilde{K}^T\sum_{l=1}^N(M_l)^Th=d^Tg=d^T(g^x+g^f+g^b)   
\end{equation}
or 
\begin{equation}\label{increment_37N}
\Delta I_{37N}\simeq \sum_{l=1}^{n_{obs}}(y_l-H_l(x^b(t)))g_l,
\end{equation}
namely, each observation contributes to the calculation of the increase, where $g\simeq \frac{1}{N}\Tilde{K}\sum_{l=1}^N(M_l)^Th$, $g^x$ contribution from initial condition increment, $g^f$ contribution from surface forcing increment, $g^b$ contribution from open boundary increment and $M_l\equiv M(t_0,t_0 +l\Delta t)$. \\
Moreover, we consider the sensitivity of $I$ to variations $\delta y_0$ in the observations, in particular, we consider $\delta y_0=d$.

\noindent Figure \ref{grafico_roms_rbl4dvar_2} shows the total number of observations from each observing platform that were assimilated into the model during cycle, increments and contribution of observations from each platform  to
\begin{itemize}
    \item NL: nonlinear model;
    \item TL: total impact;
    \item IC: initial condition;
    \item FC: surface forcing conditions;
    \item BC: open boundary conditions.
\end{itemize}

\noindent Figure \ref{Fig_obs_impact_vs_obs_sensitivity} shows curve represents the evolution of
the ocean state vector $I$ in time based on the either the prior or posterior control vector. 
In the case of the observation impact, the actual contribution of each observation, $y^o_i$, to the change $\Delta I$ defined in (\ref{formulation_increment}) (blue curve vs. black
curve) due to data assimilation is revealed. Conversely, the observation sensitivity quantifies the change $\Delta I$ that will occur in $I$ a (red curve vs. blue curve) as a result of
perturbations in the observations, $\delta  y=\textbf{d}$. In Figures \ref{forecast_impact}, \ref{forecast_2} and \ref{forecast_3} we show the results obtained by running related tests on SCoPE.

\end{itemize}

\begin{figure}

{\includegraphics[width=.5\textwidth]{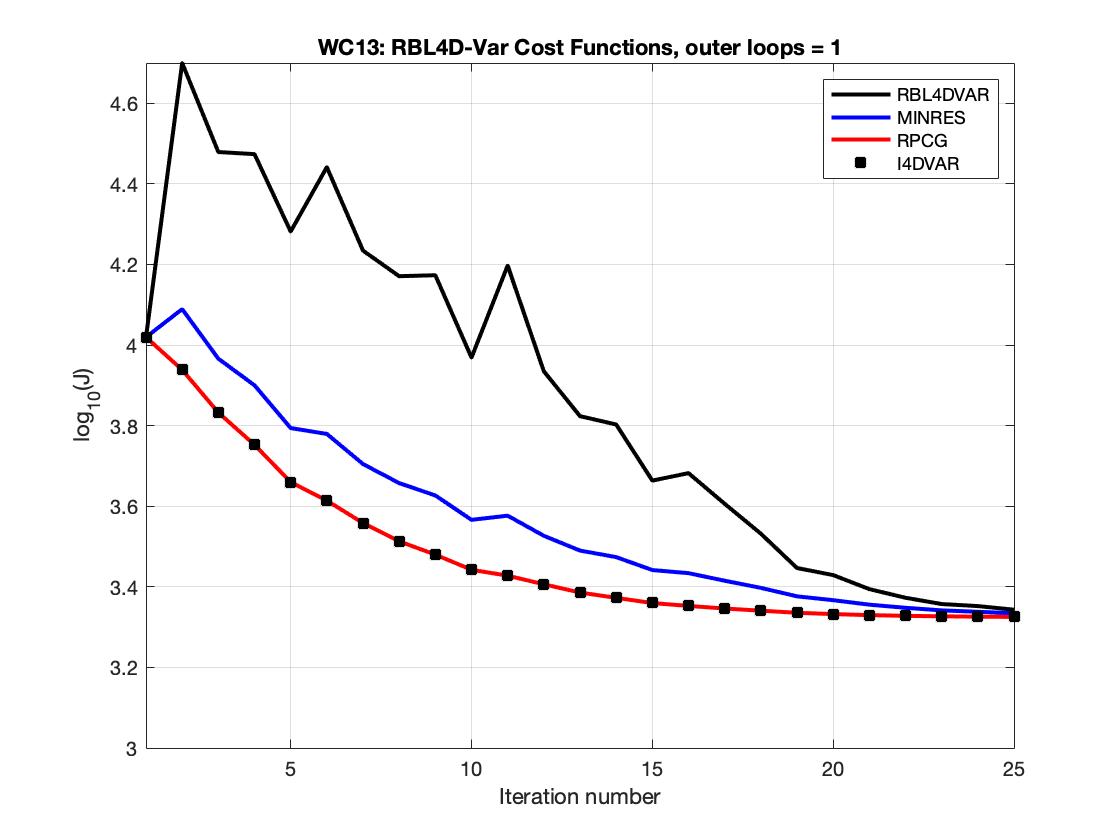}}
\quad
{\includegraphics[width=.5\textwidth]{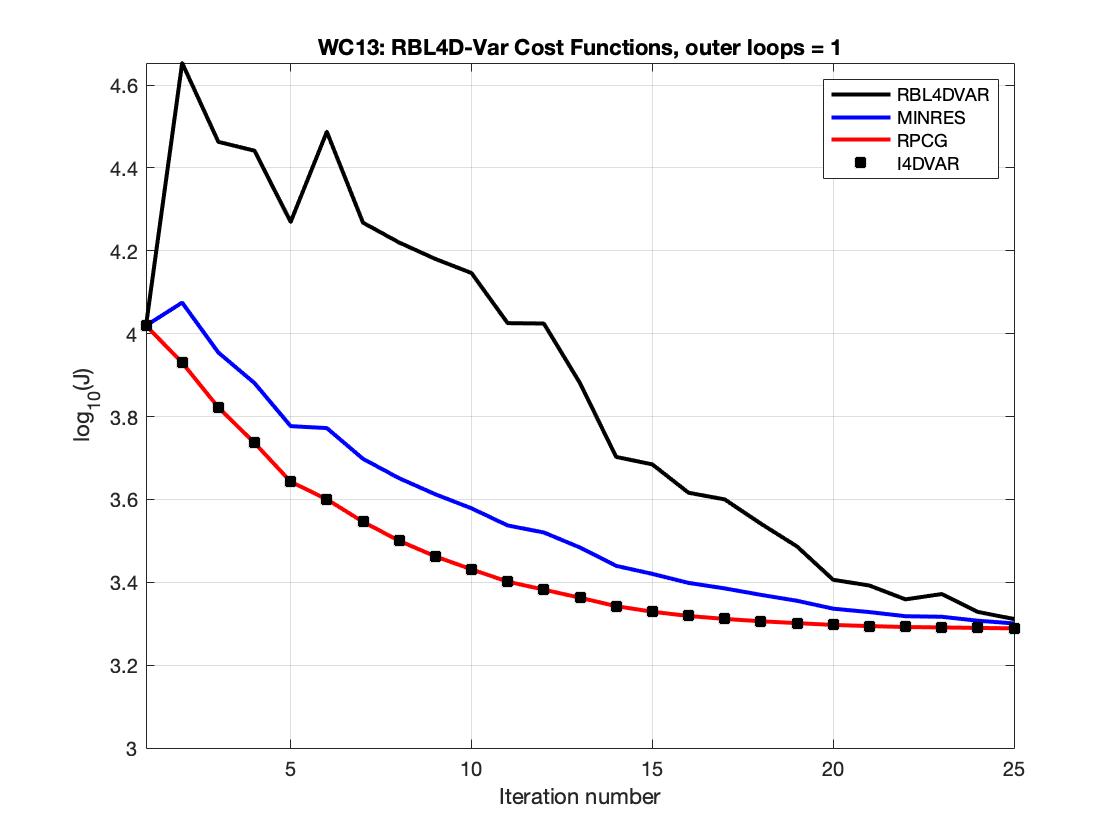}}
{\caption{RBL4D-Var Cost functions using roms\_wc13 3\_2hours.in (left) and roms\_wc13\_daily.in (right) file input, $Nouter=1$ and $Ninner=25$.}

\label{grafico_roms_rbl4dvar}}
\end{figure}

\begin{figure}
{\includegraphics[width=.5\textwidth]{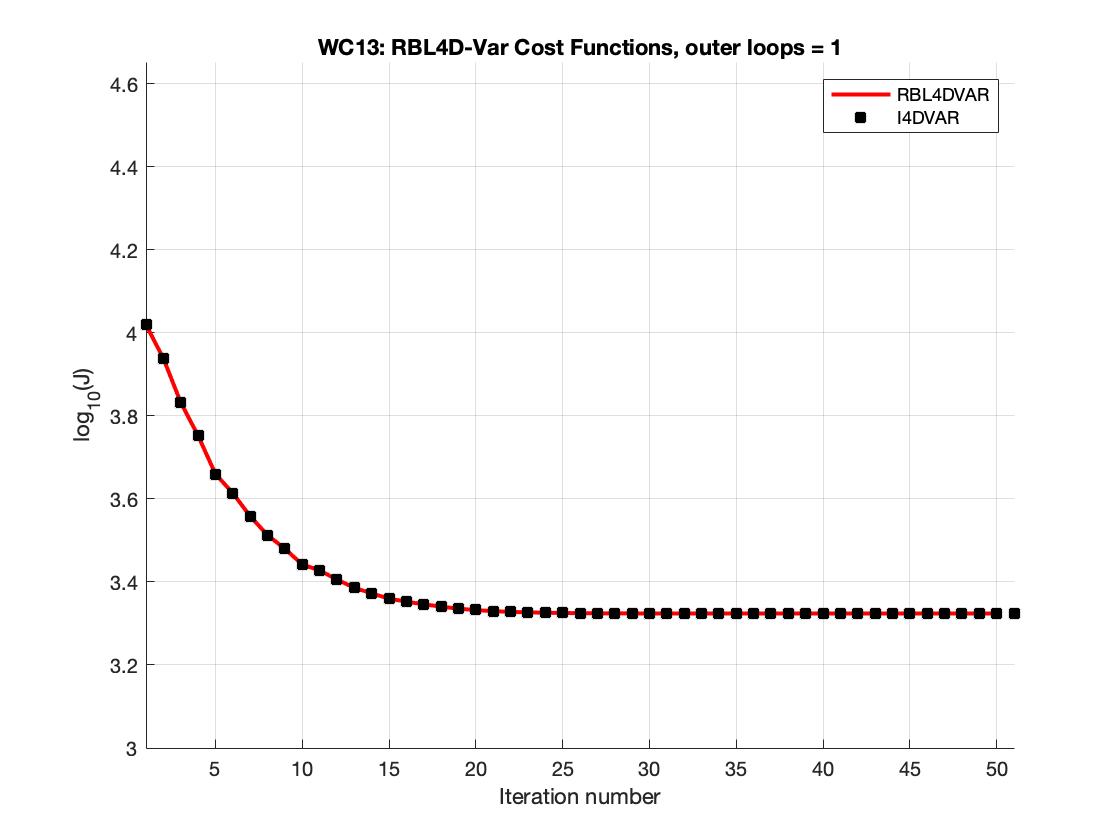}}
\quad
{\includegraphics[width=.5\textwidth]{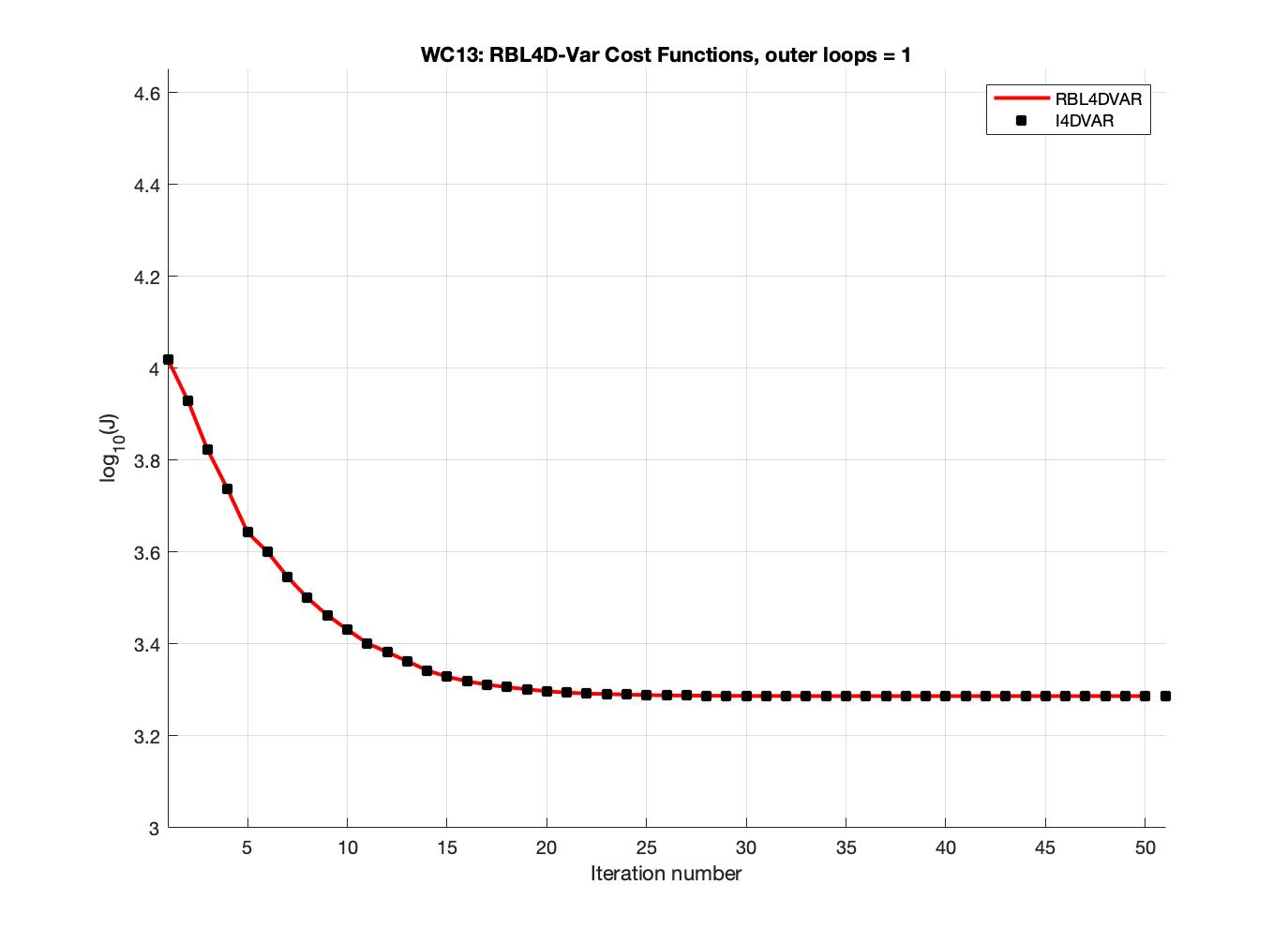}}
{\caption{RBL4D-Var Cost functions using roms\_wc13 3\_2hours.in (left) and roms\_wc13\_daily.in (right) file input, $Nouter=1$ and $Ninner=50$.}

\label{grafico_roms_rbl4dvar_2}}
\end{figure}

\begin{figure}

{\includegraphics[width=.6\textwidth]{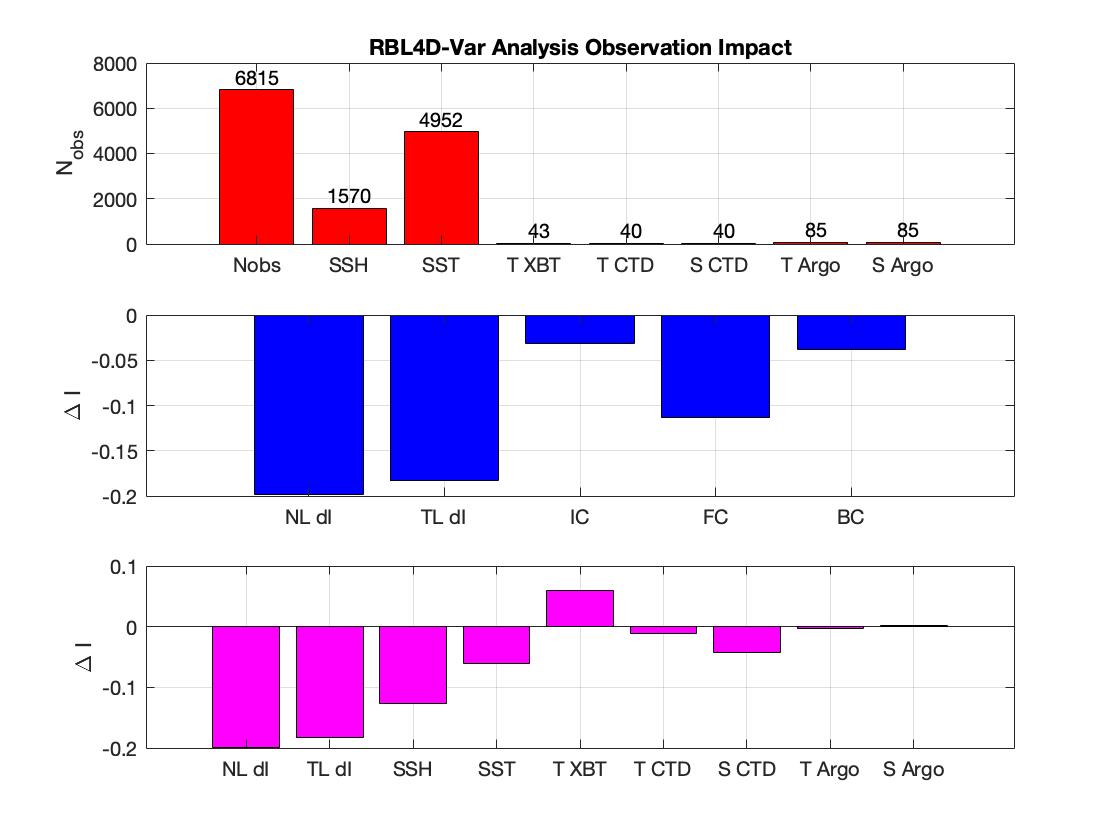}}
\quad
{\includegraphics[width=.6\textwidth]{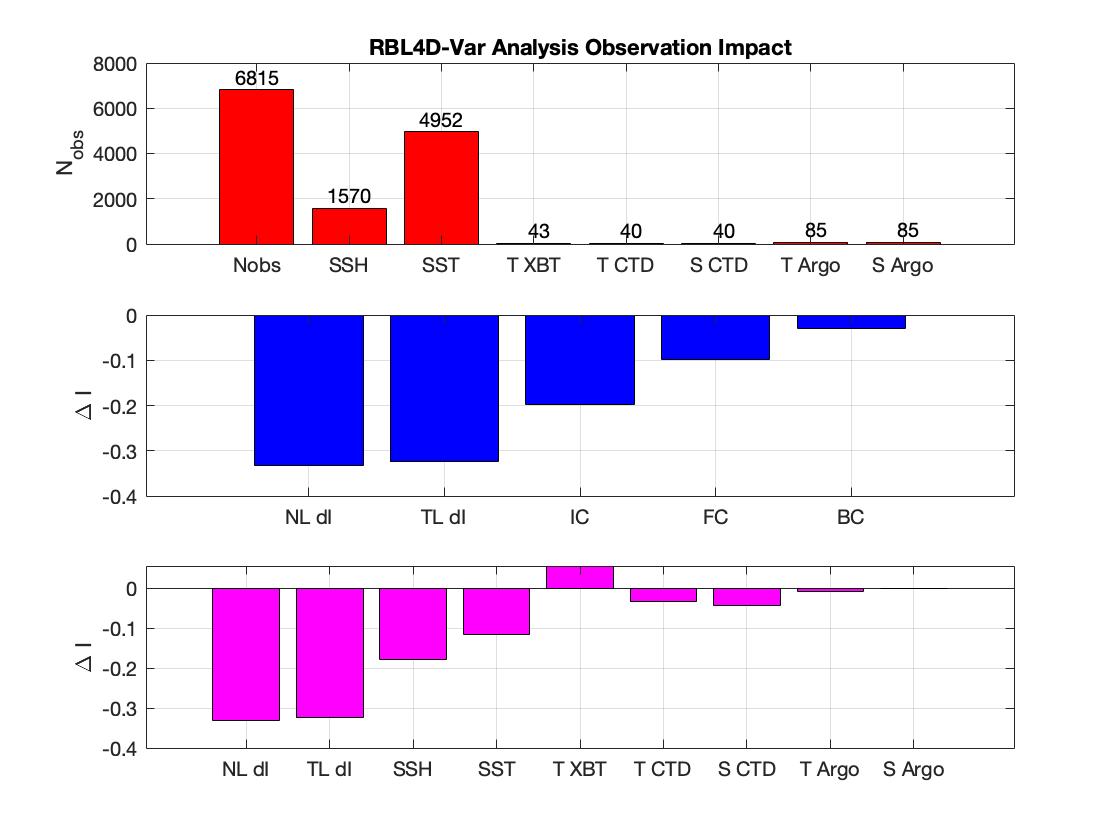}}
{\caption{RBL4D-Var observation impact using roms\_wc13\_2hours.in (left) and roms\_wc13\_daily.in (right) file.}

\label{grafico_roms_rbl4dvar_3}}
\end{figure}

\begin{figure}[!ht]
 
 \centering
  \includegraphics[width=.8\textwidth]{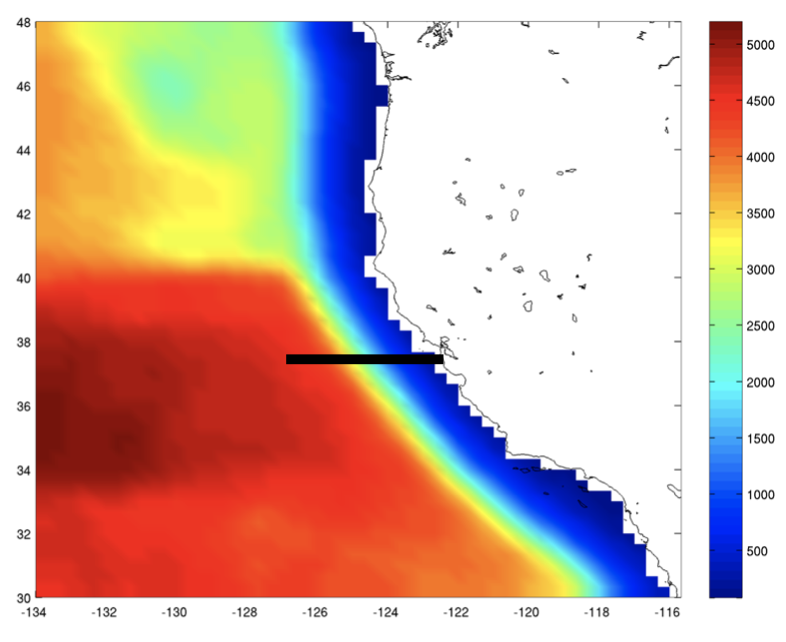}
 \caption{The 37N section along which the time averaged transport is computed from the surface to a depth of 500m.}
\label{domain_obs}
\end{figure}
\begin{figure}[!ht]

 \centering
  \includegraphics[width=.8\textwidth]{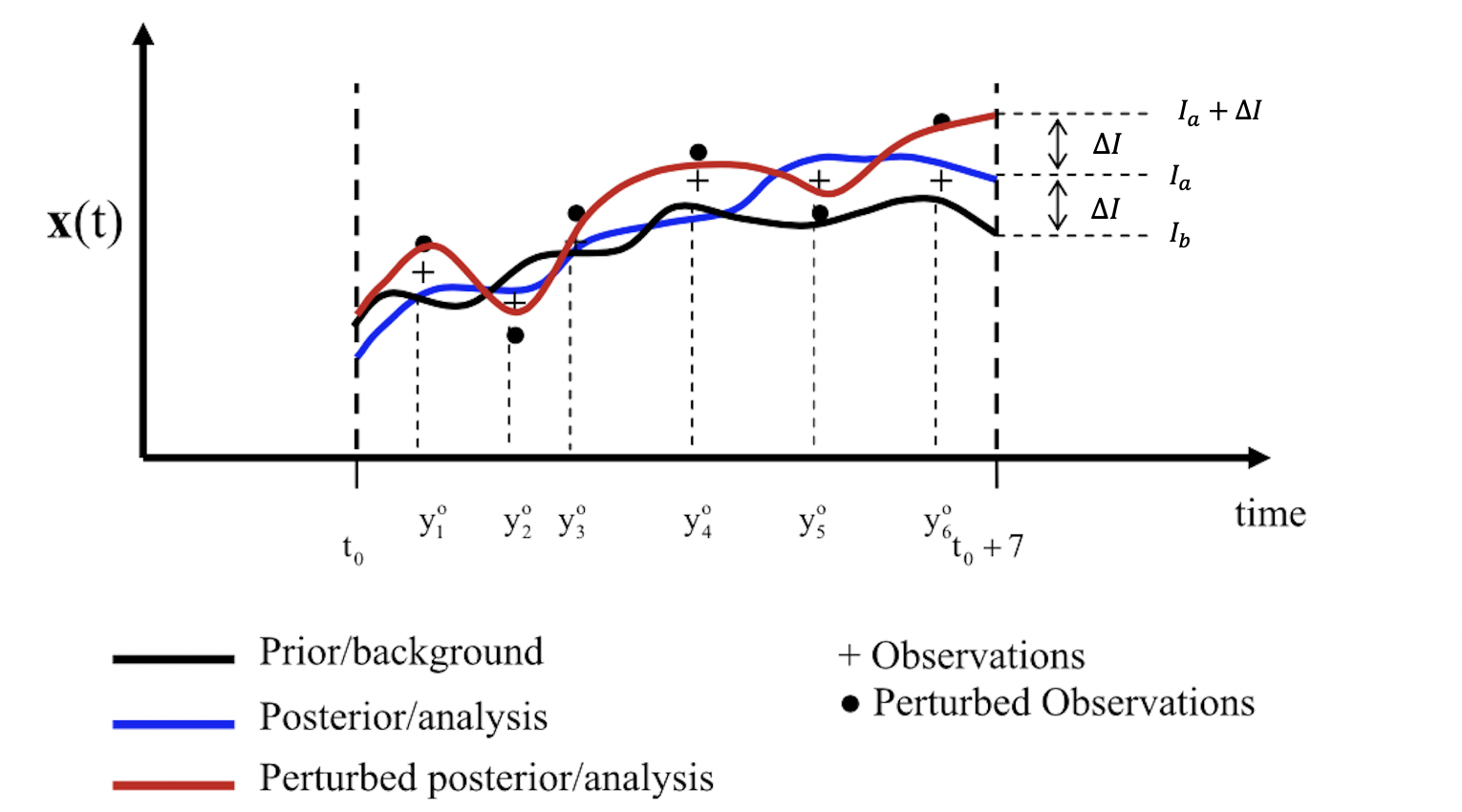}
 \caption{A schematic illustrating difference between the observation impact and observation sensitivity calculations. Fixed time interval $[t_0,t_{0+7}]$,  $y^o_i$ are observations at various times that it are shown as pluses (+) while perturbed observations, $y^o_i+\delta y^o_i$ are indicated by filled circles ($\bullet$). Consequently, red curve is perturbed analysis and blue curve is 4DVar analysis.}
\label{Fig_obs_impact_vs_obs_sensitivity}
\end{figure}

\begin{figure}

{\includegraphics[width=.6\textwidth]{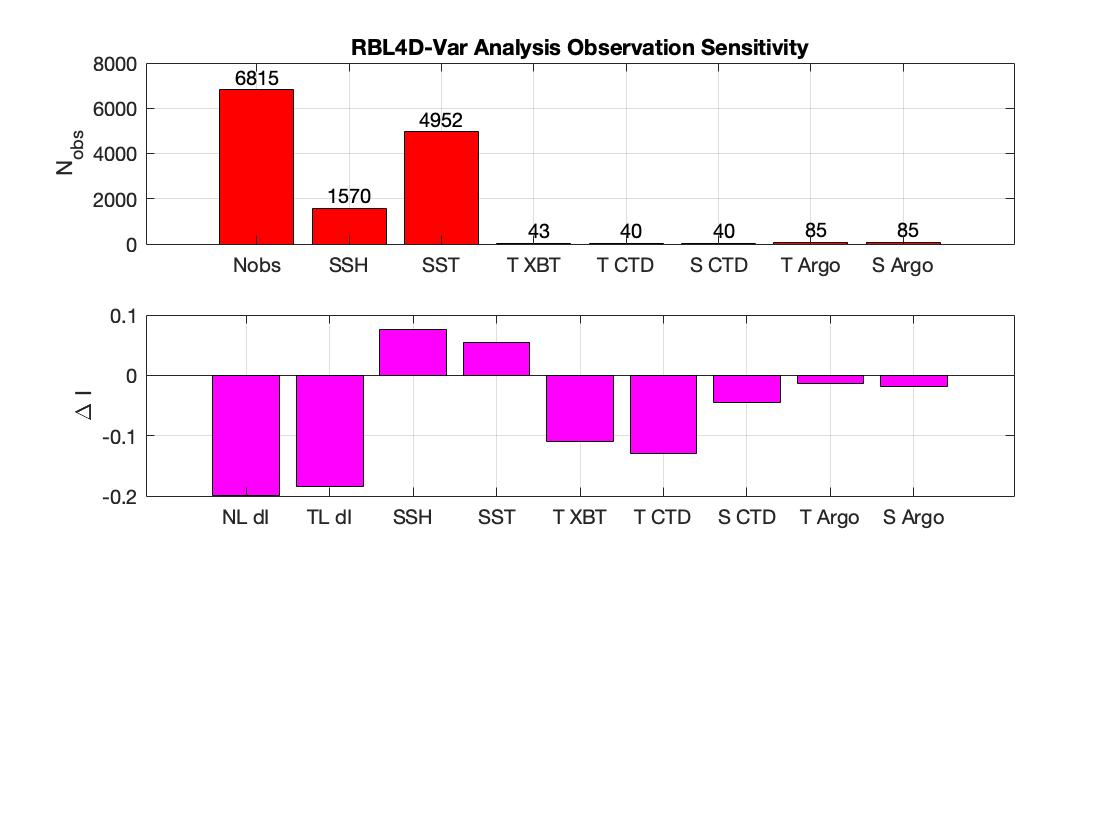}}
\quad
{\includegraphics[width=.6\textwidth]{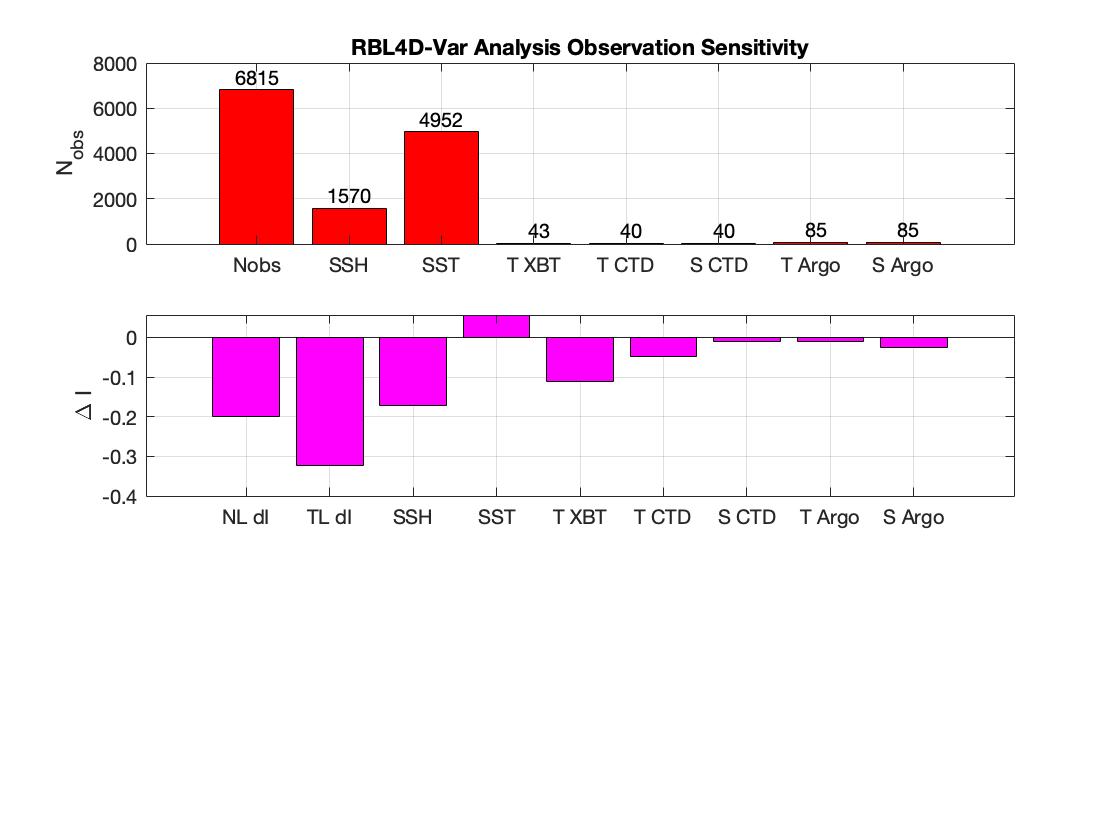}}
{\caption{RBL4D-Var observation sensitivity using roms\_wc13\_2hours.in (left) and roms\_wc13\_daily.in (right) file.}
\label{grafico_roms_sensitivity}}
\end{figure}

\begin{figure}

{\includegraphics[width=.6\textwidth]{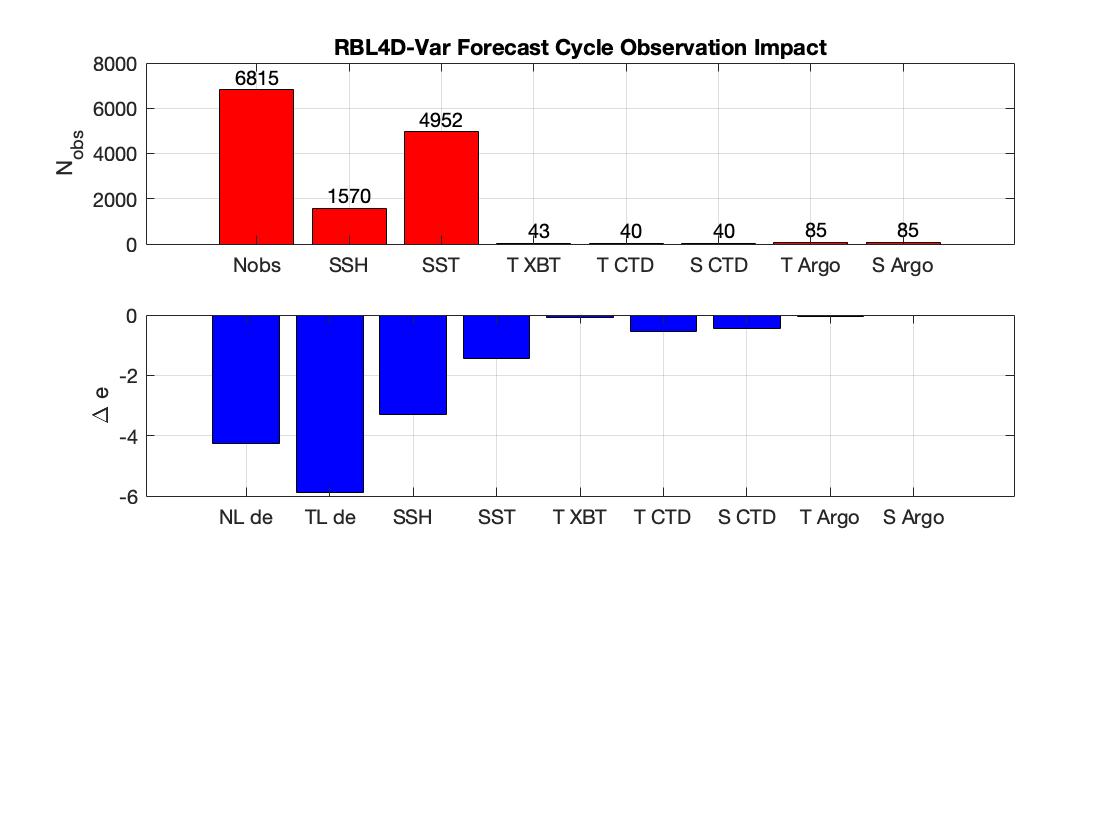}}
\quad
{\includegraphics[width=.6\textwidth]{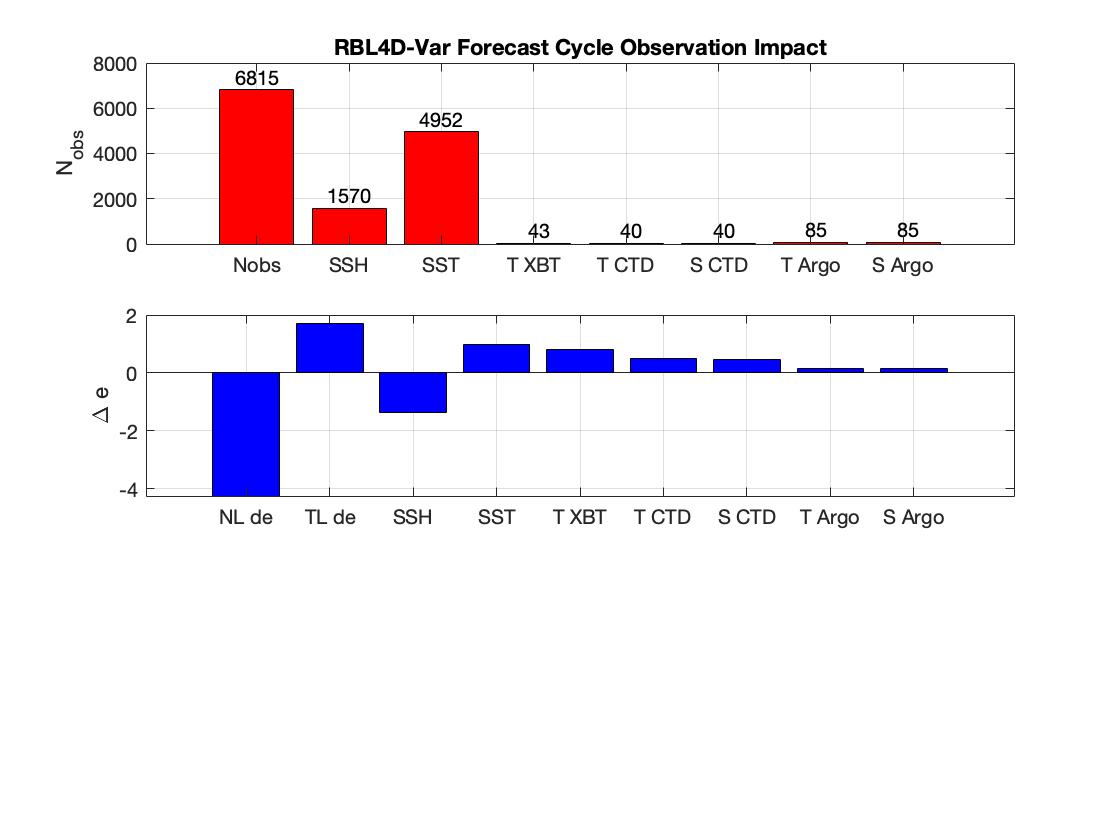}}
{\caption{RBL4D-Var forecast impact using roms\_wc13\_2hours.in (left) and roms\_wc13\_daily.in (right) file.}
\label{forecast_impact}}
\end{figure}
\begin{figure}

{\includegraphics[width=.6\textwidth]{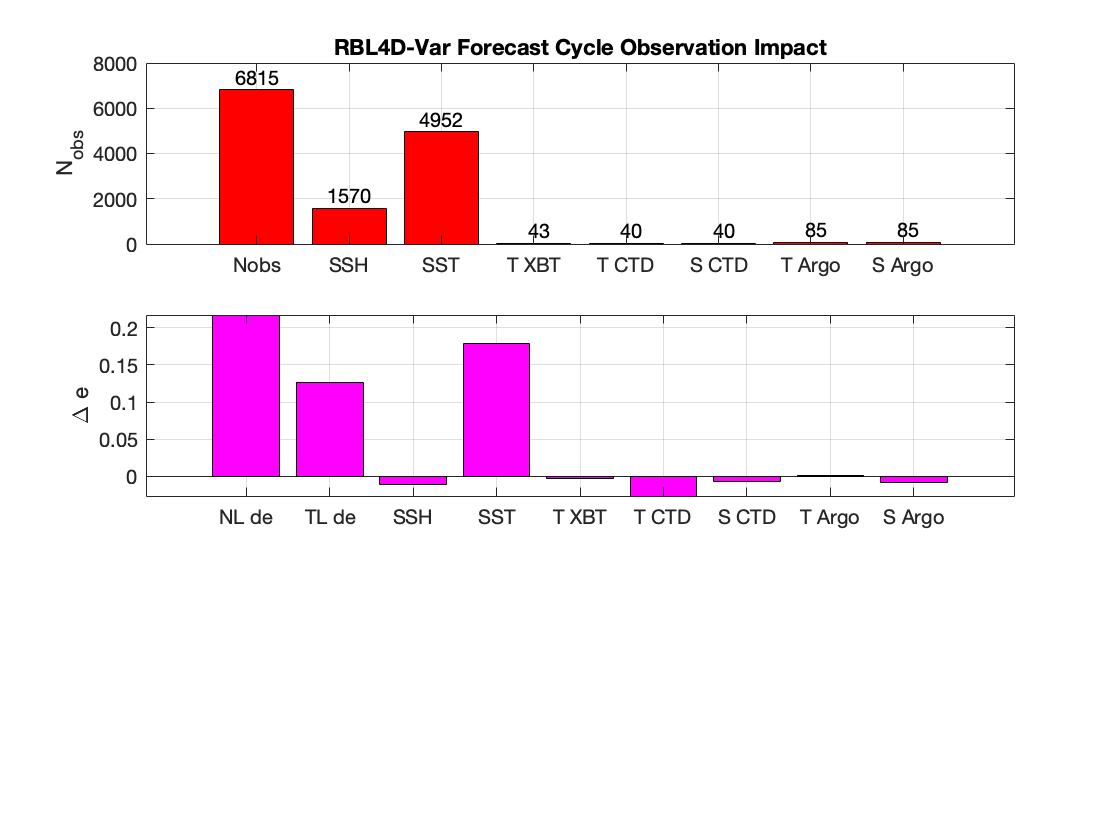}}
\quad
{\includegraphics[width=.6\textwidth]{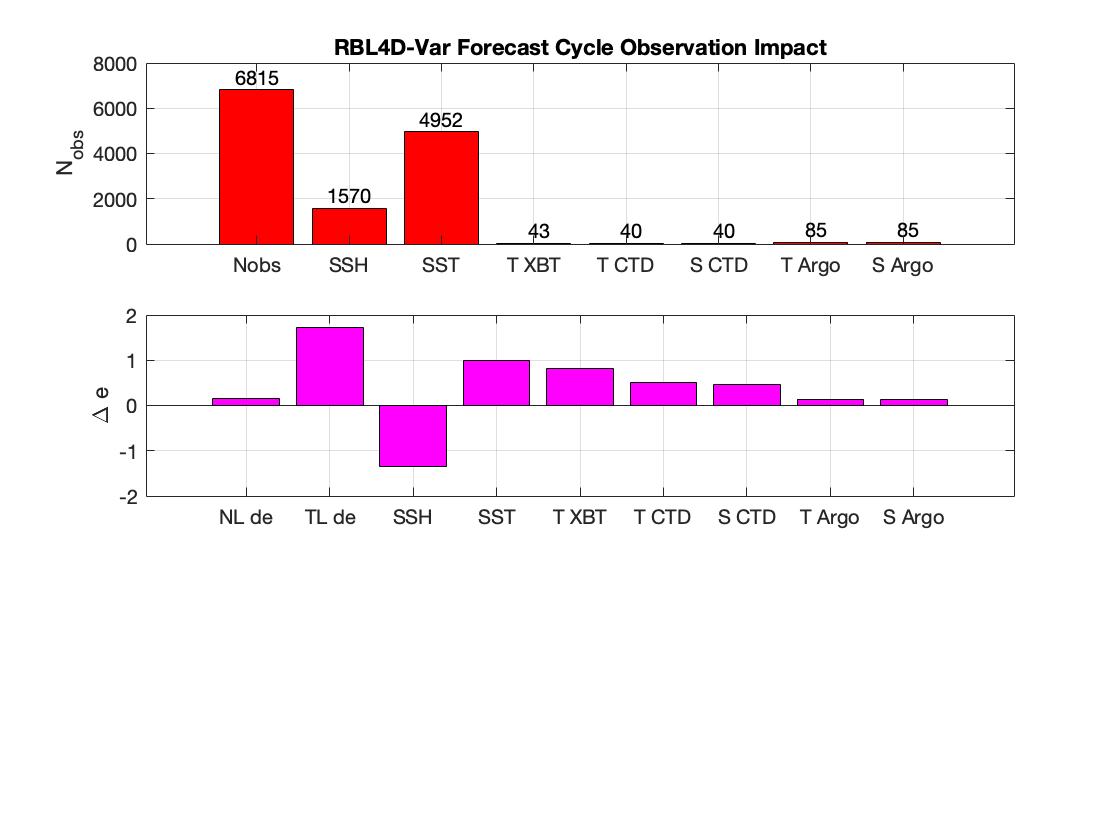}}
{\caption{RBL4D-Var forecast impact in observations space using roms\_wc13\_2hours.in (left) and roms\_wc13\_daily.in (right) file.}
\label{forecast_2}}
\end{figure}

\begin{figure}

{\includegraphics[width=.6\textwidth]{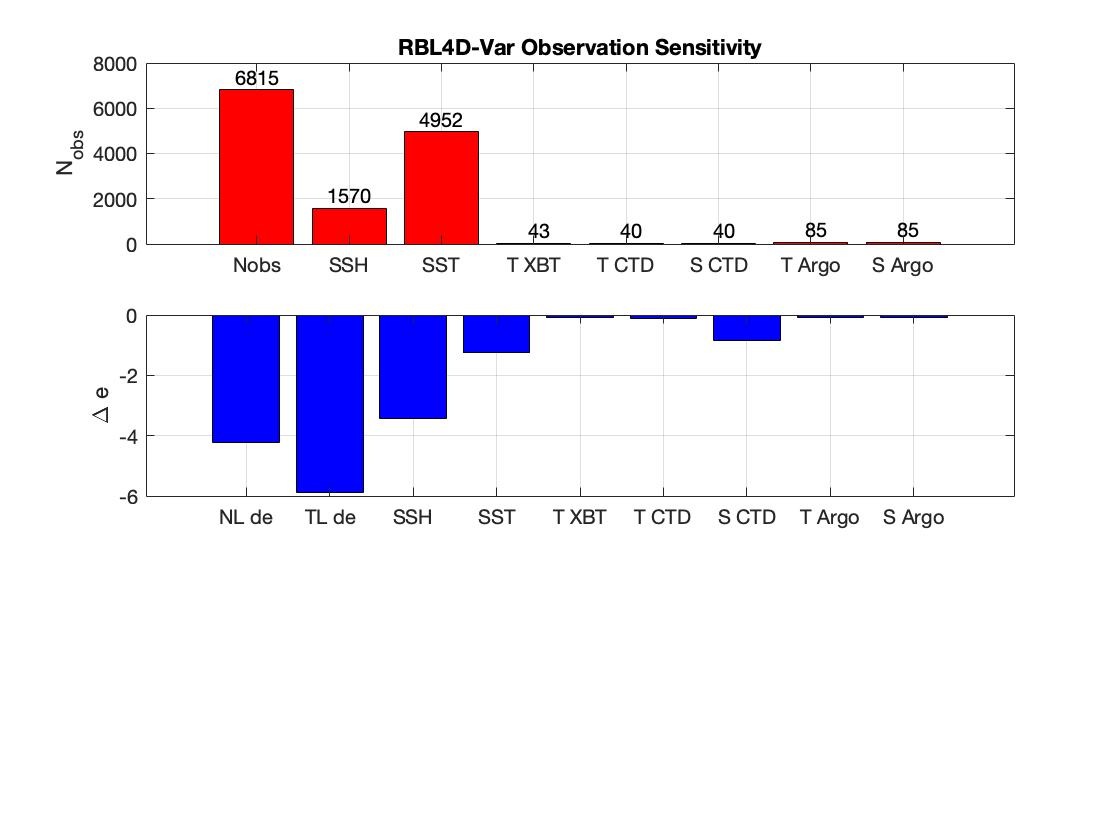}}
\quad
{\includegraphics[width=.6\textwidth]{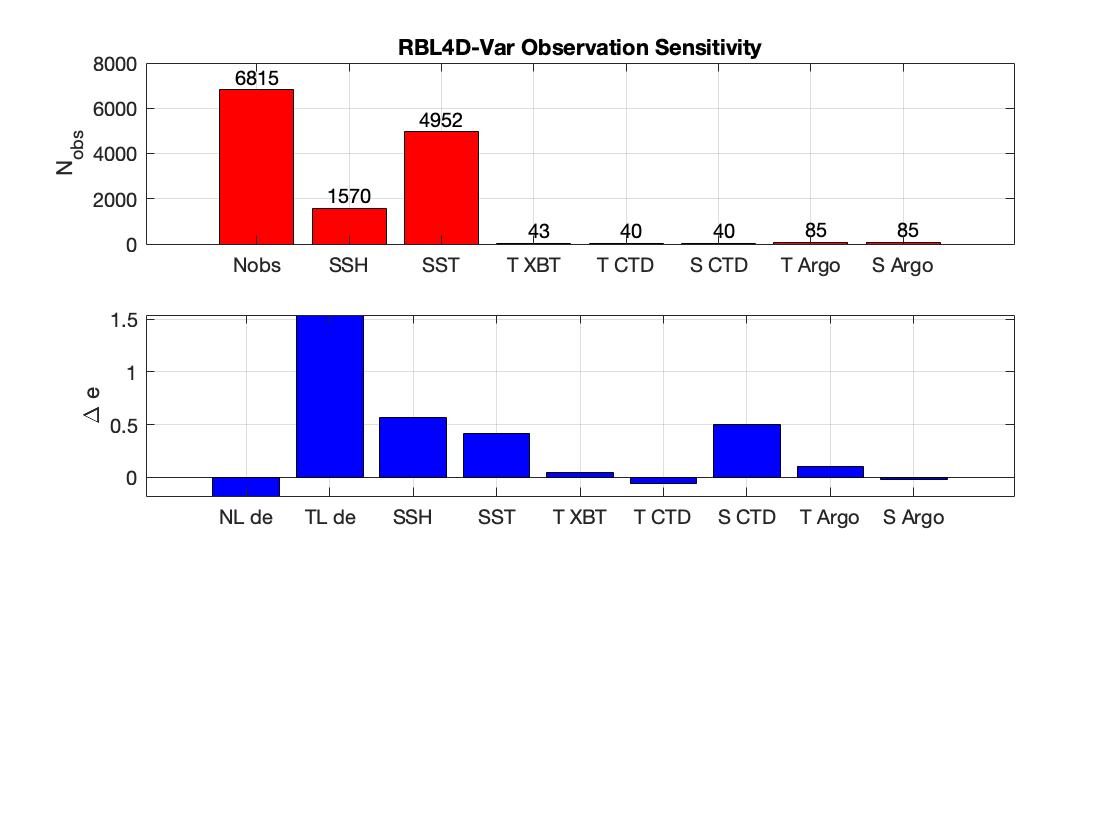}}
{\caption{RBL4D-Var forecast sensitivity using roms\_wc13\_2hours.in (left) and roms\_wc13\_daily.in (right) file.}
\label{forecast_3}}
\end{figure}

\section{Future developments}
\noindent Main modification to apply DD in time in ROMS is described by point (c) in Action 1.1, i.e. introduction of MPI communications in time. More precisely, we need to introduce and manage the MPI communication in space and time. 
\noindent The introduction of DD in time involves the following communications among processes:
\begin{itemize}
    \item Intra communications: by splitting MPI communicator (OCN\_COMM\_WORLD) to create local communicators (TASK\_COMM\_WORLD) to allow communications among processes related to same spatial subdomain but different time intervals.
    \\
    MPI commands are 
\begin{itemize}
    \item  MPI\_Comm\_split: partitions the group of MPI processes into disjoint subgroups and creates a new communicator (TASK\_COMM\_WORLD) for each subgroup. 
    \item MPI\_Isend and MPI\_Irecv: sends and receives initial conditions by setting the new communicator obtained from MPI\_Comm\_split.
\end{itemize}
    \item Inter communications: by creating new communicators (OCNi\_COMM\_WORLD) to allow communications among processes related to different spatial subdomains but same time interval. \\
    MPI commands are 
\begin{itemize}
\item MPI\_Intercomm\_create: creates an intercommunicator for each subgroup.
  \item MPI\_Isend and MPI\_Irecv: sends and receives boundary conditions by setting the new communicator obtained from MPI\_Intercomm\_create.
\end{itemize}
\end{itemize}

\end{document}